# Toward the quantification of cognition:
## *What kind of machine is a human brain?*

R.Granger *

*6207 Moore Hall, Dartmouth College, Hanover, NH 03755, United States*

A B S T R A C T

The machinery of the human brain – analog, probabilistic, embodied – can be characterized computationally, but what machinery confers what computational powers?  Any such system can be abstractly cast in terms of two computational components: a finite state machine carrying out computational steps, whether via currents, chemistry, or mechanics; plus a set of allowable memory operations, typically formulated in terms of an information store that can be read from and written to, whether via synaptic change, state transition, or recurrent activity.  Probing these mechanisms for their information content, we can capture the difference in computational power that various systems are capable of.  Most human cognitive abilities, from perception to action to memory, are shared with other species; we seek to characterize those (few) capabilities that are ubiquitously present among humans and absent from other species.  Three realms of formidable constraints --- a) measurable human cognitive abilities, b) measurable allometric anatomic brain characteristics, and c) measurable features of specific automata and formal grammars --- illustrate remarkably sharp restrictions on human abilities, unexpectedly confining human cognition to a specific class of automata ("nested stack"), which are markedly below Turing machines.

---

**Outline**

*Richard.Granger@gmail.com

**I.  Introduction: toward quantification of cognition**
What kind of machinery underlies human thought?  The terminology alone ('machinery', 'thought')
can be misleading: decades of inquiry into the nature of humans' (and other animals') mental
abilities have yielded few if any specifications that characterize such abilities, nor many exacting
measures, nor even formal lists of these abilities.

By contrast, if we wished to characterize, say, the thrust of a rocket engine, or the tensile strength of
mild steel, these are readily specified.  But to characterize the performance of, say, an artificial
language system such as Siri, there is no clear set of measurements.  Siri's *syntax* can readily be
checked by straightforward principles, but her *meanings* cannot be checked by any currently-
known principles.

This strange fact is crucial to emphasize: the only "correctness" measures of the advanced
utterances of a system such as Siri (or Alexa, or the sundry others) are in terms of empirical testing
across broad ranges of examples.  Not in terms of known principles or specifications.

Siri makes jarring errors that even a young child would not make, and it is still unclear how to
formalize them.  If I ask Siri "where is the nearest gas station?" and she replies "Dialing Stillnorth
bookstore," in what sense is that response deemed an error?  By Siri's internal logic, the response
was somehow the correct output computed from the input she received.  The sole reason that we
describe it as an error – in fact, the sole reason that we know that Siri makes errors at all – is that it
doesn't *make sense to us*; i.e., we humans do not understand how Siri's answer is connected with
our query.  This is still just empiricism: we intuit that a human wouldn't make that response, and if
they did, we could probe what misunderstanding or mis-hearing led to it.  That is starkly different
from rockets or steel: for those measures, there are principles and mathematics that structurally
organize any queries about engine or material performance.  Those metrics are separate from any
human empirical judgment call.  We may protest that those are artificial systems, whereas Siri
(although also artificial) is aimed at mimicking natural systems.  Exactly.  The lack of formal
principles or specifications for natural systems is what is at issue.  There is no a priori spec sheet
for what human language performance (for instance) is "supposed" to be.  There are right answers,
and we cannot yet say what they are.

At present this is a rigid limit to constructing systems that are intended to be at all "cognitive." For
any typical "cognitive" system such as Siri, the benchmark achievements are attained largely in the
absence of deeper understanding, unlike the case for systems such as engines and materials.

Marr (Marr 1982) presciently warned of the need to carefully particularize the "computation" or
"specification" level of a task, that is, the characterization of its overall abilities, independent of the
candidate mechanisms (implementation and algorithms) that may produce those abilities.  A
proposed system built for a given task cannot succeed if that task is ill-defined.

A software system could be constructed with the intention to carry out, say, handwriting
recognition; and extensive effort, testing, and validation steps could be undertaken, only to
eventually discover that the task itself had been ill-defined, and thus seemingly-accurate
recognition systems may abruptly encounter instances on which they radically fail, such as
"Captchas."

Until the appearance of captchas, handwriting recognition software seemed swimmingly successful,
but this was an illusion: the software succeeded on extant datasets, but the field did not actually have an adequate characterization of handwriting. Humans recognize simple captchas effortlessly, apparently using the same perceptual processing that they employ in recognizing other, less noisy, handwriting. But the candidate systems that were being forwarded for handwriting recognition fell apart entirely when presented with Captchas.

*Richard.Granger@gmail.com

It is worth emphasizing that, until the appearance of these problematic captchas, the field solidly believed that it was succeeding (LeCun et al. 1998). There was no widespread acknowledgement that the field was studying a task that would very soon utterly fail to match human performance on these simple extensions to handwriting – extensions that humans themselves had no trouble with. Captchas caught the field unawares.

From its apparent and illusory successes, the field abruptly fell flat. The field took roughly another *ten years* to finally achieve reliable Captcha recognition (George et al. 2017).

We are in a similar epoch today. A predominant focus on the statistics of huge data (such as games, image classification, shopping recommendations) artfully alters the metric for success. Rather than addressing open-ended problems as humans typically do, they aim at achieving a known metric (e.g., a game win). These games are being won by massive memorization and processing of millions of instances; see, e.g., the meticulous recent review by Serre (Serre 2019). This is highly reminiscent of the decades-old standard computer science field of numerical computation, which may have no important relation to intelligence or cognition, let alone human intelligence or cognition – but is highly successful at 'big data'.

AlphaGo has played millions of games: i.e., the equivalent of the combined life experience of an imagined hundreds of thousands of Go players. Humans perform highly complex tasks of recognition, retrieval, decision (not to mention creativity and invention) in the absence of million-sized datasets.

This many-orders-of-magnitude difference in amassed experience is sufficiently vast that it can reasonably be characterized as a difference in kind. A human playing Go is not performing the task that AlphaGo is performing.

But, but, they both make the same moves! They both can be measured by their wins and losses! Surely they are indeed doing the same task! What is that task? If it is measured purely behaviorally, as pieces moving on a board, the tasks are the same. If there is even the remotest interest in the mental behaviors of a player, then the specifications may be wildly, incommensurately distinct.

Lacking these specifications of actual human mental behavior, it can be all too easy for us to imagine that we are formally addressing a task. Just as the field of handwriting recognition indeed imagined, for many years before captchas. Task specification is the crucial starting ingredient that matters; otherwise we get million-data-memorizers such as the AlphaGo/AlphaZero/AlphaStar family.

The lesson is not being learned. Tasks are carefully steered away from far-reaching human abilities, focusing instead on data memorization mixed with slight generalization, as in game playing. By this nostrum, a stream of attractive and lucrative successes are toted up, largely disregarding the failures and shortfalls (Serre 2019).

In fact, the same approach that gave the appearance of succeeding so impressively at initial handwriting recognition, and then failed so spectacularly when faced with Captchas, is still one of the dominant approaches in the field.

To be sure, these sedulous approaches deserve acclaim for their grueling analyses of colossal data sets --- but we may ask what are the next Captchas that will starkly illuminate the shortcomings of current approaches?   Most common-sense knowledge; most language semantics; most of our everyday understanding of the real world.  Today's "intelligent" systems can "read" books, and then prove unable to answer some of the most basic questions about what happened in the books. Marcus and Davis (Marcus and Davis 2019) provide illustrative examples: after the "Google Talk to Books" system read the Harry Potter books, when asked "How did Harry Potter meet Hermione Granger?", no answers came close to answering the question, and most were wildly off-point.

(Many in the field are very mindful of the situation, aware that most current approaches provide measures of just the narrow efficacy of specific programs on given families of datasets.  Researchers are actively striving for applicable metrics that could assess systems more broadly, potentially identifying "IQ-like" scalar measures, especially for proposed artificial general intelligence (AGI) systems.  There are a range of views on this approach; see, e.g., (Legg and Hutter 2007; Hernandez-Orallo, Dowe, and Hernandez-Lloreda 2014; Dowe and Hernandez-Orallo 2012; Chollet 2019). These approaches are seeking "feasible and reasonable" tests to which a system could be subjected, probing "universal" abilities rather than "customized" for particular tasks (Hernandez-Orallo, Dowe, and Hernandez-Lloreda 2014).  A separate question is this: what kind of machine generates the abilities that they are measuring?  Studies of that kind appear to target any constructed subset of universal Turing machines (artificial or natural), without distinguishing among the *differential* abilities that arise at different sites within the automata hierarchy.)

This problem is not about language or reading per se.  It is about the representation and manipulation of knowledge.  The sentences in the books are largely parsed correctly, and semantics of individual statements are coded.  However, human readers educe huge additional amounts of information from such sentences.  "John saw Jenny's name in the paper" implies that John had a newspaper (whether ink or screen), and held it, and read it, and that there was a story in it that discussed a person named Jenny, and that John evidently knew something about this Jenny independent of the newspaper story.  Even seemingly simple sentences have this property: they signal large amounts of real-world episodic knowledge that is represented in the reader's head (e.g., (Frankland and Greene 2014) for relevant studies).

The organization and manipulation of knowledge is a crucial part of cognition.  It underpins language understanding, and much more: seeing sequences of interactions in a movie or play, without dialogue, can readily convey correspondingly large amounts of knowledge in an observer.

Learning and memory – the acquisition and storage of that knowledge – is the central difference between biological intelligences and the current set of artificial supposed approaches to intelligence.

When the term "representation" is used in the artificial neural network literature, it typically is expropriated to denote graded levels of feature combination in successive backpropagation layers. This of course specifically lacks a crucial element in wider usage of "representation" throughout the cognitive literature: that of relations.  Mental representations include relations among entities (above, before, within); among episodic content (such as explicit temporal sequence information); and more.  Distinctions are crucial if we wish to carefully characterize the differences among cognitive systems. So-called "representations" that merely conflate features into simple composite entities within "isa" hierarchies are a highly restricted subset of more general relational representations.

To address this distinction, many researchers attempt to construct "hybrid" systems that use artificial neural networks to analyze perceptual information on one hand, and symbolic or rule-based systems for higher conceptual or relational information, on the other. But in a brain, if there are both "low-level" neural operations and "high-level" symbols, presumably the latter arise directly from the former, and the two interact seamlessly. "Hybrid" systems avoid addressing this fundamental seamlessness of information in brains. Previous studies have shown how the physiological operation of thalamocortical anatomic circuitry can be directly interpreted in terms of grammars, which in turn can embody rules; these studies and related work may aid in identifying the underlying links between low- and high-level processing in human brains (Rodriguez, Whitson, and Granger 2004; Granger 2006; Rodriguez and Granger 2016).

Until we can characterize the nature of our stored information, and the nature of the brain circuit machinery that stores, retrieves, and manipulates it, we have not understood the fundamentals of human cognition. Possibly, as current machine learning would have it, all our stored knowledge is nothing more than stockpiled statistics educed from big data. The simple examples of John and Jenny, and Harry and Hermione, and many, many more, argue otherwise. It appears that current artificial systems are not adequately characterizing crucial features of human cognition.

With these admonitions in mind, we pursue specifications (Marr's "computation" level) for human cognitive capacities. The pursuit entails multiple threads across disciplinary boundaries that may initially appear unrelated. Of particular note are i) the anatomical regularities of brain allometry, ii) the close apparent relations among multiple uniquely-human cognitive abilities; and iii) the equivalence classes of formal automata. We show how these may be related to each other to engender a formulation of the computational capacity of human brains. That is, what kind of machine is a human brain?

## II. Summary: behavior, allometry, automata
a) *Qualitative and quantitative characterizations of empirical human cognitive behaviors*
Humans are plainly different from other primates: we drive cars, read books, build houses. Yet clear specifications of the abilities that characterize us have proven maddeningly elusive. Other animals make war, transmit culture, use tools, exhibit some number sense, infer other animals' intentions, and much more; ongoing work attempts to distinguish uniquely human forms of such abilities from nonhuman (possibly antecedent) versions; e.g., (Herrmann et al. 2007; Suddendorf 2013).

Abilities such as counting, logic, perspective-taking, all may share some characteristics across species, as examples of "triune" capacities: those for which there appear to be three separable variants. Number sense, for instance, has a variant typically referred to as "numerosity" (Brannon and Terrace 1998), an ability to distinguish some quantities (e.g., two vs five) without being able to count arbitrary quantities (Gelman and Gallistel 1978). In addition to the nonhuman (numerosity) and human (counting) variants, there also are much more advanced variants (arithmetic, calculus, etc.) possessed by many humans but very far from universal.

These three variants: rudimentary, uniquely human, and advanced, may be broad categories applicable to multiple human capabilities. The "rudimentary" versions of abilities (such as numerosity) may occur in nonhumans or in humans during early development. In most cases, the precise relationship between nonhuman capabilities, on one hand, and the capabilities of humans during early development, on the other hand, remains unknown; the abilities are not yet sufficiently quantified to enable detailed comparisons. Similarly, the highly advanced versions of uniquely human abilities are divergent and variable (arithmetic, logic, calculus, linear algebra, ...); it remains unknown (and controversial) whether the formal computational capacities underlying all such abilities (some core "math ability") are related or distinct.

Without resolving these further subdivisions, we adopt a triune terminology for cognitive capacities, distinguishing between those that either are simpler than, or equal to, or surpass, unique human abilities. The triune distinction intrinsically generates a class of abilities that are neither rudimentary nor expert: that is, they are unique to humans and not other species, and are exhibited by all humans (below the abilities of highly trained experts). The resulting central class is that of "all and only" abilities: those possessed by every human (barring neuropathology) and all humans (not just trained experts).

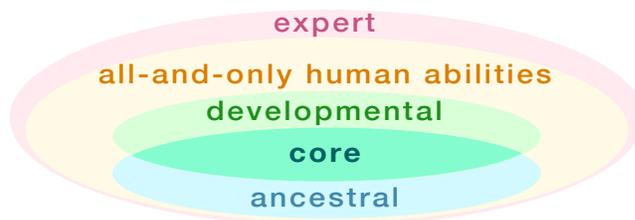

Figure 1. Some abilities ("expert") are only acquired via training plus external apparatus (e.g., paper and pencil); some ("ancestral") are present in us and also in other animals; some ("developmental") occur in primitive early form before maturing. Some are present in all humans and no other animals ("all-and-only"); these abilities presumably have some combination of "core" (developmental and/or ancestral) precursors.

We posit that these "all and only" human abilities are highly worthy of study in their own right, as well as in contrast to their possible rudimentary and expert forms. In particular, if a given ability is possessed by approximately every human but evidence indicates that it is not possessed by nonhumans, then something in us endows us with this ability. Throughout, we strive to focus on distinctions among the three clearly distinguishable categories of abilities – non- or early-human; all-and-only human; and expert-human – as they arise.

*b) Brains: Allometric comparative anatomy*
Although evolution is alleged to produce "endless forms most beautiful" in Darwin's poetic phrase, actual biological product is severely constrained. All mammals have nearly identical limbs, toes, organs; famously, giraffes have the same number of neck vertebrae as mice and mammoths.

Brains are no exception: all are astoundingly similar across four orders of magnitude in size in mammals. The size and connectivity of almost every brain component and connection pathway can be computed with disturbing accuracy, from just the overall size of a brain (Finlay and Darlington 1995; Striedter 2005; Herculano-Houzel et al. 2007; Lynch and Granger 2008) via a simple set of allometric equations. In other words, all mammalian brains obey the same severely constrained set of rigorously predictable designs.

(Note the essential distinction between *relative* size and *allometrically-predicted* size. Relative sizes of brain parts change drastically, as in the case of area 10, yet they do so entirely predictably, as a function of straightforward allometric equations associated with each area – equations that account for more than 99% of the size variance of all brain areas, across all the hominids, including humans (Finlay and Darlington 1995; Semendeferi et al. 2002) (see Figure 2).)

In sum, the parts of a primate brain, and the way those parts are wired together, are rigidly determined by allometric relations.

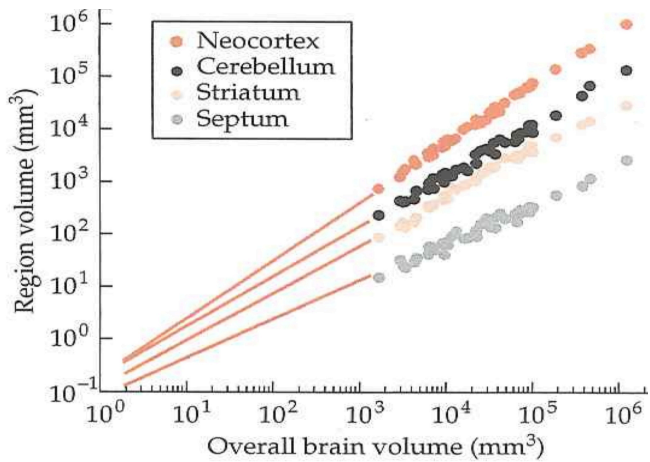

Figure 2. Distinct allometric slopes per brain region. Slope determines the volume percentage allocated to that brain area, as overall brain size (x axis) varies. If the slope is >1, the area becomes a disproportionately larger percentage of the brain, as brain size increases. Phylogenetically earlier structures have lower slopes; relatively later structures (including neocerebellum) have higher slopes. See also Figure 6.

The rigidity of allometry should shock us. Where is selection? Where, for that matter, are all of Darwin's "endless forms"? Divergences from allometry do exist – so-called "mosaic" species differences that can indeed be selected for. They are surprisingly rare, and, when they occur, surprisingly minute (e.g., the small divergences from regression lines in the figure).

Notable exceptions occur in nonhuman animals, yet the search for specialized brain 'organs' in humans – ala a cow's stomach, or bat's radar – thus far come up surprisingly short. Human brains are exquisitely consistent with sheer allometry, giving as yet no new hint of how a given observed difference could mechanistically enable gifts such as human language to spring forth in us and in no others; see, e.g., (Yang et al. 2017; Tattersall 2017; Hauser et al. 2014; Berwick and Chomsky 2016; Fedorenko 2014; Friederici et al. 2006).

If allometric relations so rigidly obtain, then many of Darwin's "possible" forms are evidently not being generated. It is *not* that they are first generated and then selected against. There are no living beings nor fossil records of substantial divergences from allometry. Something in our genome prevents them from being generated, or from successfully making it through development, in the first place. This is starkly different from selection, in the sense of organisms being born with differences, and competing in environments, and possibly dying off. Allometry does not indicate any of those steps: rather, genomes operate in such a way that they adhere to allometric constraints throughout development and adulthood, as opposed to trying out different forms and letting them "compete". These facts about allometry are widely discussed in the evolutionary literature; see, e.g., (Wagner 2014). Many, many instances of "evolved" traits (such as human area 10 size) exhibit no evident characteristics of selection, variation, or evolutionary pressure. They are allometrically predetermined (Striedter 2005; Semendeferi et al. 2002). The change in *overall* brain size may be considered an evolved trait; but any additional novel brain attributes (such as sizes of particular areas) arrive as preëstablished parts of the package of allometrically concerted changes; i.e., traits that are forcibly, allometrically, yoked to the change in the overall brain size.

Human brains are unexpectedly large compared to our body sizes – but human brains are no exception whatsoever to the allometric relations of all other primates (and mammal) brains. Our brains rigidly adhere to the same equations. The divergences between human and other primate brains are breathtakingly tiny. To extraordinary accuracy, human brains are precisely scaled-up chimp brains (Herculano-Houzel 2009, 2012). These stringent constraints are among the most powerful fundamental biological phenomena underlying any candidate hypotheses of human brains.

This, of course, begs the question of how we, alone among the primates, live in houses, drive cars, administer vaccines. We are at least as social as other highly-social primates; we uniquely cook food; we uniquely read other people's intentions; we uniquely use arithmetic. Why only humans? What kind of machine is a human brain, that it confers these unique talents?

*c) Computation: Specification of automata and grammars*
Pioneers of computability theory (Turing, Church, Hilbert, Kleene, Gödel, Rosser, Post, and others) introduced formal properties of computation, their limits (such as undecidable problems), and their equivalences. The Church-Turing thesis, for instance, identified the equivalence of Turing's abstract computing device (an automaton termed an "*a-machine*"), $\lambda$-calculus (a formal logic system), and recursively enumerable functions (a specification of a language, i.e., a formal grammar) (Church 1936; Doyle 2002).

Each machine (automaton) can be seen as a finite representation of a (potentially infinite) formal language (set of acceptable sequential "strings"), and automata are typically classified according to the specific set of languages they can "recognize," i.e., the grammars that their computations can produce or parse.

Chomsky, Post, Aho, Hopcroft, Ullman, and others (Chomsky 1975; Hopcroft and Ullman 1969) provided a containment hierarchy, classifying automata and their associated grammars by these computational powers, from least (finite state machines, computing regular grammars) to most powerful (Turing machines, computing recursively enumerable grammars).

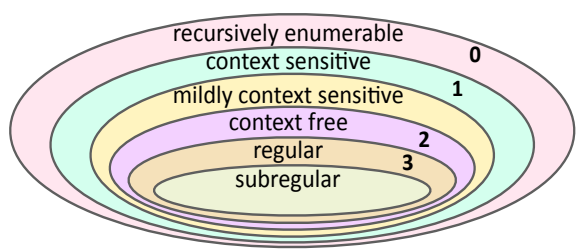

Figure 3. A version of the grammar hierarchy (see Fig 12 for comparison). The simplest finite state automata have no stack memories, generating subregular and regular languages. Addition of unrestricted single stack memory yields context-free and mildly-context-sensitive languages. Expansion to multiple, independently addressable stacks of restricted length generates context sensitive languages. Removing the stack length constraint produces recursively enumerable languages. (The original hierarchy had Type 0,1,2,3 grammars, shown.)

Notably, the automata/grammar hierarchy establishes equivalence classes. That is, seemingly different machines can be found to be equivalent in terms of what they can accomplish, i.e., what they can compute, as in the discussion of Marr in the Introduction. (See also Figure 12). We can build a mechanical adding machine out of gears, levers and other parts --- and we can build an electronic adding machine out of transistors, diodes, and other components.  The two may have identical output behaviors: present two numbers and their sum will be produced (possibly with the same latencies).  If so, the two machines are equivalent in terms of their "computation" level, but they differ in the steps that they carry out (algorithms) to arrive at those ultimate computational outputs, and they differ further in the materials used to embody those algorithmic steps (i.e., their "implementations").

The distinction among these levels (computation, algorithm, implementation) establishes "equivalence classes" at the computation level, i.e., a range of different machines that nonetheless exhibit directly comparable task-level behaviors.

If these equivalence classes can be identified for particular tasks, that will not mean that the brain mechanisms for those tasks will have been discovered.  It may be that brains use different implementation methods, or even different algorithms, to carry out the equivalent computations. Nonetheless, equivalence classes may be illuminating.  At present, existing systems such as Alpha Go are not being directly compared to careful characterizations of human Go-playing behaviors.  As long as the aim is simply to numerically win the game, little is being learned about human intelligence at all.  As improved characterizations of human behavior and cognition are developed, our understanding is correspondingly incremented.

Turing's initial aim was to identify the formal quantitative power of a "computer," which at that time was a clearly defined term referring to a human (or room full of humans) with paper and pencil, following instructions.  For large projects (such as war efforts), groups of trained people ("computers") were recruited to sit at tables and carefully obey instructions to carry out written manipulations per instructions from group leaders.  (The eventual term "*electronic* computers" was, of course, by contrasting reference to the original (human) computers.)

Armed with arbitrary access to pads of paper (and calculating machines) from which they could read out information and to which they could record information, those carefully-instructed humans could achieve large-scale mathematical calculations --- operations beyond what individuals were able to attain on their own, with no external appurtenances.

That pattern is captured and formalized in one of the standard definitions of a Turing machine, which consists of a module to carry out instructions (a "finite state machine", or FSM), and a defined way to write to, and read from, a specified set of memory storage locations (a "stack", or a "tape"). The formal automata hierarchy defines multiple distinct levels of computational power, each denoted by its FSM plus some specific connate memory system, denoted by specific formal terms, from so-called "restricted" (R) stack memories, through "unrestricted" (UR), "nested" (N), "bounded" (B), and "unbounded (UB) stack systems.  (The terms can be misleadingly counterintuitive compared to their non-mathematical dictionary definitions; for instance, "unrestricted single stacks" are far weaker than "bounded multiple stacks."

Cognition is not directly carried out by machines in the automata hierarchy; there is no intention to conflate the two.  Rather, the different levels of complexity of various mammalian brains, from mouse to mammoth, may be evaluated for their measurable cognitive abilities.  There has been evidence offered of automata-cognitive relations, in rodents, in primates, in humans, and even in avian species (see, e.g., (Fitch and Friederici 2012; Hagoort 2019; Fitch 2014; Petkov and Jarvis 2012)).

*d) Summary*
Our focus is the quantification of cognition: the ability to measure magnitudes of human mental capacities and to test their correspondence to known metrics of artificial automata. (Such an approach stands in contrast to uncalibrated methods such as i.q. tests, or other simple compilations of scores on tasks.) At the time of Church, Turing, Post, and others, no such metric data existed. Even during the later era of Chomsky, Hopcroft, Ullman, Aho, and colleagues, there still were no such measures expressing formally quantified evaluations of any complex human capacities. Beginning in the 1980s, and continuing to this day, there are numerous quantitative assessments, predominantly in the realm of specific human language capacities. Remarkably, their findings all are in close agreement with each other (Vijay-Shanker and Weir 1994; Joshi, Vijay-Shanker, and Weir 1991; Weir 1992).

Extant quantitative evaluations of human cognitive limits, together with the well-studied allometric constraints on brain structure, severely constrict the set of formal automata that conform with the data. The resulting findings were unexpected, yet they are solidly grounded; they bespeak a seemingly inescapable assessment of the computational power that is intrinsic to human minds.

## III. All humans; only humans.
As introduced above, we address empirical specifications of human brains and cognition, applying the findings to the automata and grammar hierarchy. In the following subsections, we introduce a small fixed set of critical constraints, in the three domains of:
   - cognition: empirically observed species-specific mental capabilities,
   - circuitry: comparative neuroanatomy across species, and
   - computation: situating behaviors and biologies among automata and grammars.

Together, the intersection of these aspects of neuroscience, automata theory, linguistics, comparative anatomy, comparative ethology, and brain evolution, are shown to establish a staggeringly restrictive set of constraints about the intrinsic capacity of human brains. We formulate and discuss what they entail for admissible candidate propositions about human cognitive capabilities.

### A. Qualitative and quantitative characterizations of empirical human cognitive capacities
*Thesis A.1) Humans are qualitatively different from other primates*
There used to be a "human uniqueness" list: human abilities that definitively distinguished us from other organisms. On the list, at various times, were reasoning, war, culture, deception, tool use, and many more. We have recognized that other organisms exhibit all of these behaviors; crows plan, and use tools, chimps war exceedingly well, many animals deceive, and many pass down cultural knowledge, skills, and social organization. The "unique to human" list has grown perplexingly brief.

There are multiple behaviors, however, for which there appear to be human and nonhuman "versions"; human behaviors for which there may be nonhuman analogs or precursors. Wherever such antecedent capabilities occur, the distinctions may be studied distinguishing a possibly-unique human version from nonhuman versions.

These human-nonhuman pairs of behaviors include the inference powers of chimps, as potential antecedents of human formal logic (e.g., (Cesana-Arlotti et al. 2018; Hanus et al. 2011; De Waal 2016)); "numerosity," possibly partially underlying human formal mathematics (Brannon and Terrace 1998; MacLean et al. 2012); various calls, vocalizations, signals and other communication procedures that may be partial precursors to human language (Berwick et al. 2011; Beckers et al.

2012).  There also are phenomena of cultural learning and social interaction in nonhumans, that may prefigure more complex human-specific social machinations (Herrmann et al. 2007; MacLean et al. 2012; Suddendorf and Corbalis 2010; Suddendorf 2013); and other abilities appearing in nonhumans but having more advanced ostensible counterparts in us.  Pinker & Jackendoff (Pinker and Jackendoff 2004) particularly note the value of language, and its utility for other candidate all-and-only human characteristics, such as "a reliance on acquired know-how and a high degree of cooperation among non-kin" (Pinker and Jackendoff 2004; Tooby and DeVore 1987).  Suddendorf (Suddendorf 2013) lists a set of suggestive terms for candidate pairs of these nonhuman–human cognitive counterparts: communication → language; memory → mental time travel; social reasoning → mindreading; physical reasoning → theories; empathy → morality; tradition → culture.

In each case, not only are there nonhuman antecedents to human abilities; there also are developmental phases in human infants, before a full uniquely-human ability has appeared, during which developing humans appear to exhibit the antecedent, in ways that may be identical to or closely corresponding to the nonhuman version.

*Thesis A.1a)  Some humans can do things that the majority of humans cannot*
The editor of this journal likes to remind his students (and colleagues) that, as far as we know, calculus was only discovered by (at most) three human beings.  Everyone else had it taught to them.

We all live in houses and drive cars and use laptops.  A tiny number of those reading this are capable of building a house, car, or laptop.  Or even a bottle, or a pencil.  We may not be able to readily conceive of a world in which we had to invent almost all we use, or discover almost all we know.  That does not resemble our world.  It does resemble the world of non-human animals.

One human discovered (invented) the Turing machine.  A few others made highly related similar discoveries (e.g., Church, Hilbert, Post, Kleene).  All the rest of us came to understand Turing machines directly via instruction stemming from those initial few, a drastically simpler achievement.  (An intriguing, but separable, question is whether no others at the time might have made these timely discoveries if Turing, Church, and others had not existed).

The difference between discovery and instruction is whether or not someone handed you the fully-formed set of information.  At the core of that difference is the deceptively simple fact that information can be transmitted.  Doing so ineluctably requires a medium with specific characteristics: it must be able to be recorded on and read from.  It might be designated as an external, non-volatile read-write memory system.  Put simply: writing things down.  Multiply two nine-digit numbers?  All but impossible in your head; trivially simple on paper.  The distance between these is decisive.

As discussed in detail in a later section, the sole distinction among different automata are their memory systems.  The only thing that separates them are their exact capabilities for storage and retrieval.  A Turing machine **is** a finite state machine, plus a specific set of added storage and retrieval mechanisms.

Much more on this later.  For the moment, the takeaway is this: a human with a tape recorder (or pencil and paper or chalk or tablet or laptop or camera) is not at all the same thing as a human lacking any access to these.  In our own everyday lives, these are so inextricably entangled with our thoughts and behaviors that we may give them little credit.  Take away the storage tape, and the Turing machine is not a Turing machine: it is a finite state machine, with a vastly different scale of computational power. Correspondingly, take away our external storage devices, from recorders and laptops down to paper and pencil, and our computing power may be enormously diminished. We posit that commensurate limitations hold for biological computing engines as for artificial ones.

In both cases, the fundamental question at the heart of computational measurement is the one asked at the beginning of this article: what machinery confers what powers?

*Thesis A.2) There is a compact set of human ubiquitous and unique (all-and-only) capabilities*
Most human abilities do not arise inborn, like a foal's ability to walk; rather, they each appear to emerge at roughly prescribed latencies, in putative correspondence with developmental stages: evidence clearly distinguishes different ages at which children exhibit specific concepts of numeric capacities, mental time travel, theory of mind, and others.

During the evolution of the primates, it might have been the case that different cognitive advances arose in different taxa. One could posit a species with, for instance, human-level mathematics, but lacking mental time-travel, advanced cultural learning, and morality. Or a species with logical inference and advanced social mindreading, lacking any advanced numerical abilities, morality, and natural language.

Surprisingly, evidence suggests that none of these abilities occur independently in different species; they all evidently arise simultaneously in a single species. Our recent relatives include a set of possible ancestors whose constituency continues to grow rapidly as new discoveries become sequenced. These include Neanderthals; Denisovans (Krause et al. 2010) and other possible paleo Siberians; recently reported Homo luzonensis (Détroit et al. 2019). Since all of these other than Neanderthals have been discovered within the past few years, it is not unreasonable to imagine that other recent hominins may be found. It is notable in this regard that the vast majority of "anatomically modern human" fossils have not been sequenced. If there were species who exhibited subsets of the above list of "uniquely human" abilities, they are extinct and their mental abilities currently are impossible to ascertain. Homo sapiens – the sole known species that has *any* of the "human" versions of these capacities – has *all* of them. Quite possibly, all of these abilities arose phylogenetically at once, in us.

These circumstances intimate a possible link among these abilities, rather than coincidental coöccurrence. The question of dependencies among these abilities is unanswered. To what extent do any of these abilities depend on other abilities – is there some ur-ability that is both necessary and sufficient for the emergence of this plethora of seemingly-disparate capacities?

A2a) Numerosity
"Numerosity," a reduced version of numeric abilities, appear to arrive relatively early (e.g., 4 to 6 month old children, notably before onset of uniquely human language abilities. By contrast, the strict human ability to count (which also entails at least learning the words for numbers), typically does not appear before roughly age 6; Gelman & Gallistel (Gelman and Gallistel 1978) proposed principles that counting is dependent on, including "abstractness," meaning any objects can be counted, "cardinality," i.e., the ability to arrive at an announced number (such as "5": 1,2,3,4,5), "order irrelevance," meaning the ability to start with any of the objects being counted (not just, e.g., the leftmost object), and "one-to-one" assignment, i.e., any given object is counted only once, wherever it occurs in the sequence.

Correspondingly, development of logic was assumed by Piaget (Piaget 1952) to be intrinsically linked with that of number: "a pre-numerical period corresponds to a pre-logical level," and "logical and arithmetical operations therefore constitute a single system that is psychologically natural, the second resulting from a generalisation and fusion of the first...". Infants' "numerosity" abilities have been cited as evidence against this view, but such an argument may be seen as conflating the nonhuman version of number sense (numerosity) with strictly human arithmetic abilities that do not arise until far later in development. There is evidence that by roughly age 5, children have acquired relatively simple arithmetic principles such as "complementarity": if adding two and four results in six, then taking two away from six must equal four. Before that age, evidence indicates that children as old as 3 to 5 still do not have this concept (Starkey and Gelman 1982; Starkey, Spelke, and Gelman 1990). Very interestingly, 19-month infants looked longer when an object A surprisingly appeared where B was thought to be (Cesana-Arlotti et al. 2018); 12-month-olds exhibited concordant (albeit less statistically significant) behavior; these may be evidence of logic (specifically, modus tollendo ponens), although simpler hypotheses remain consistent with the data (Jasbi et al. 2019).

A2b) Theory of mind; perspective-taking

Knowledge of what another agent knows and intends ("theory of mind") similarly appears to have nonhuman and human versions. Chimpanzees may be able to infer whether other chimpanzees do or do not know the whereabouts of some hidden object (Hare, Call, and Tomasello 2001) whereas chimps evidently fail to infer whether or not a human knows such a location(Povinelli, Nelson, and Boysen 1990). And again, correspondingly, humans develop these abilities over time, from early (7-9 months) knowledge of whether someone is paying attention (Baron-Cohen 1991), to the eventual ability to recognize different desires in different others, and to the understanding that a person may have false beliefs (Sue thinks the toy is in the box, but Bill removed it when Sue wasn't looking). (And encompassing these theory-of-mind abilities, the further ability of perspective-taking enables a human to perceive or conceptualize what the environment or situation looks like from that other person's mental point of view.)

It is not known whether abilities such as language and theory-of-mind develop together just coincidentally, or whether one depends on the other (Milligan, Astington, and Dack 2007; Miller 2006).

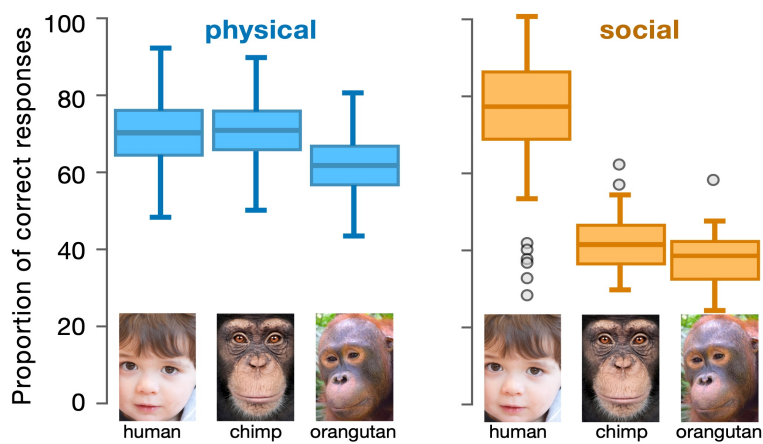

Figure 4. Evidence of quantitative cognitive distinctions. Human and nonhuman primates were tested on physical tasks (e.g., locating an object, discriminating quantities, etc.) and "social" tasks (observing and using a conspecific's solution; communicating a location; following a gaze to a target). Species were not significantly different on physical tasks but differed on "social" tasks ... once "outliers" are omitted (gray dots). (Adapted from Herrmann et al '07.)

A2c) Mental time travel
Humans are able to conceive and report on both past events and future events, including those that did not take place or will not take place (e.g., counterfactuals such as "what would the U.S. be like if the Nazis had won," or "which college will I choose if I'm admitted to both").

As with numerosity and theory of mind, there are primate precursors of this ability (Osvath, Raby, and Clayton 2010; Cheke and Clayton 2010; Suddendorf and Corbalis 2010). And, as with numerosity and theory of mind, there a developmental trajectory through which children pass before exhibiting the full ability by roughly age 3-5 (Suddendorf and Redshaw 2013; Redshaw and Suddendorf 2016).

A2d) Seriality
Human language, counting, and other abilities are serial by definition: event "b" (word or number) cannot occur until after event "a".

For some human abilities, such as comprehending a conspecific's intentionality, the quality of seriality is not at all obvious; whether or not it is a feature of this ability is an open question. (As mentioned, nonhuman primates do not have the same mind-reading abilities as humans, and human children go through an early stage where they appear to conspicuously lack this mind-reading ability. This suggests the possibility that some later capacity is required for the full ability. That later capacity, such as some form of logical inference, may entail seriality.)

Given the colossal parallelism of brains, the inherent seriality of language, counting, and other human behaviors may appear unexpected or even paradoxical. Although most computation in brain circuits is carried out in parallel, all such computations are internal. All input to and output from these computations is necessarily serial. These distinctions may bear some correspondence with others in the literature such as "fast" and "slow" behaviors(Kahneman 2011), i.e., pre-linguistic, intuitive thought, versus behaviors that call on some form of (serial) logic or language.

Syntactic structure in language is structured and hierarchical. Yet any information in such structures must also be rendered into serial sequence in order for it to be recognized or produced: we notoriously can only say, or hear, one word at a time. Seriality is in some ways so obvious and pervasive (all logic, all language) that it may be readily overlooked.

A2e) Working memory
"Working memory" is variously described as a "limited capacity" system for temporarily holding information, in which new input causes older items to be dropped; these held memories are crucial for decision making in real time. Working memory typically refers not just to temporary storage of information, but also active on-line manipulation of that information (Goldman-Rakic 1995); it designates maintenance of information about an event for some time, in the absence of reinforcement, that can be used to guide behavior at a later time (Miller, Galanter, and Pribram 1960; Baddeley 1992).

As will be discussed in detail in the following sections, the frontal cortical regions, which are consistently tied with working memory (Goldman-Rakic 1988), expand proportionately far more than any other brain structures. Whereas a human brain is routinely characterized as being roughly 3.5 times larger than the brain of a nonhuman ape of the same body size, the frontal cortex of that same human brain is roughly seven times larger than the ape's corresponding frontal cortex. It might be expected that functions linked with frontal cortex may exhibit commensurate increases.

A plethora of definitions of "working memory" make it difficult to consistently characterize, let alone quantify.  A recent metastudy (Lind, Enquist, and Ghirlanda 2015) indicated no obvious correlation between working memory duration and relative brain size or encephalization quotient (see also (Carruthers 2013) for further review).  In particular, primates did not exhibit longer working memories than rats.  However, a range of relatively recent studies provide evidence that humans exhibit a substantially larger working memory when semantic relations are introduced (Brady, Störmer, and Alvarez 2016), possibly akin to "chunking" (Cowan, Chen, and Rouder 2004; Laird, Rosenbloom, and Newell 1984); such studies question the applicability of standard testing regimens for working memory.

A2f) Are the all-and-only human abilities interrelated?
Human language is strictly species-specific.  And (barring brain disorders) every human has it: language is acquired with little to no training, merely exposure, requiring no evident effort or strain.  This is dramatically different from language training attempts in nonhumans, which are vastly effortful and as yet have yielded at best highly questionable results, drastically separating human language from animal communication systems.

This effortlessly-acquired human language capability sets us drastically above other animals.  Even more advanced abilities, equally unique to humans, such as logic or math, or even writing, are utterly unlike language in that a huge percentage of humans will never learn them, and those who do will require enormous training, sustained effort, and irreplaceable reliance on external aids such as blackboards, recordings, textbooks.  (Multiplying two seven digit numbers is trivial with pencil and paper, and highly nontrivial without.  The differences are of primary interest.  Distinctions of course matter very much if we wish to understand the nature of unique and ubiquitous human abilities.)  It is intriguing to consider that advanced math may be as different from language, as language is from other animal communication; the struggles of an ape learning language may be comparable to those of a student learning calculus.

As described, multiple other capacities besides language – such as the human theory of mind ability, human cultural transmission, human mental time travel, human perspective-taking, human moral rules – may be similarly categorized if they, too, arise in all of us, with negligible training or externalities.

This seemingly intrinsic level of human ability – the all-and-only tasks – may arise from a proposed core set of abilities.  It may be that no single one of these tasks (e.g., language) comprises the essence of that core; rather, multiple seemingly-disparate abilities may arise from the capacities of that core.

Grammatically separate statements such as "the child hit the car" and "the car hit the child" evoke notably distinct images.  Evidence suggests that the resulting syntactically-composited concepts engage very late-appearing telencephalic brain areas (such as anterior temporal cortex); these may be distinguishing among semantic roles arising from the events, whether linguistically or visually presented.  That is, expanded cortical regions are recruited to process the roles arising in complex grammatical formulations; see, e.g., (Frankland and Greene 2014).

Evidence indicates that successive cortical regions are selectively recruited to process longer and longer patterns, whether auditory, linguistic, or video (e.g., (Griffiths et al. 1998; Hasson et al. 2008; Farbood et al. 2015)), and some findings have linked hierarchical language syntactic structure to a sequential hierarchy of cortical time scales that correspond to grammar construction (Ding et al.

2015).  These all may be seen as instances of the general principle that successive cortical regions process incrementally expanded grammar rewrite-rule expressions.

The key questions raised are these:
   - what quantitative measures exist for each proposed all-and-only human capacity?
   - what dependencies may obtain among these capacities: are some contingent on others?
   - in general, what are the inclusion criteria for identifying all-and-only capacities in humans?

If we want to know what type of machine we intrinsically are, then we may ask: what are the most powerful abilities that are bestowed on all humans and only humans?  For any sufficiently powerful "all-and-only" ability, we may wish to identify what kind of underlying computational mechanism suffices to produce it.

(We may also ask a reciprocal question: if some specific computational power (e.g., Turing-equivalent) is claimed for human cognition, what exactly are the capacities that would be conferred uniquely and ubiquitously on humans by those computational powers?  In what ways can we measure their manifestations in human behavior and cognition?)

Many of these all-and-only abilities are as yet exceedingly difficult to accurately quantify.  It is notably difficult to fashion a quantitative y-axis for theory of mind, for cultural transmission, for perspective-taking.  In contrast, there is a large literature of quantitative measures of human language abilities.

*Thesis A.3) Human natural language characteristics have been consistently and replicably quantified*
As already noted, although there are extensive literatures discussing correspondences and distinctions among human versus nonhuman versions of cognitive capabilities, such distinctions are (unsurprisingly) predominantly qualitative.

In this respect, the literature on human natural language properties is an anomaly: the attributes of natural language syntax, an all-and-only human capacity, has repeatedly been quantitatively characterized.

Beginning in the 1980s, several groups of researchers, from multiple different institutions, independently engaged in ventures aimed at quantitatively appraising the formal characteristics of human natural language.  These investigations proceeded by taking corpora of various languages, and analyzing the types of syntactic transformations that occur and do not occur, i.e., that are accepted or "not accepted" in the language.  This is not proscriptive, i.e., these are not judgments of what "ought" to be accepted, but rather what native speakers judge (with surprising ease and consensus) as actually, intrinsically, acceptable versus unacceptable.  ("Who do you wonder whether they will come"; "John is very tall, doesn't he?").  Notably, these judgments also readily accept syntactically well-formed but semantically nonsensical statements; "Colorless green ideas slept furiously" is (famously) quite readable, and even graphically evocative, whether or not it makes any semantic sense.

These analyses of language size span Ph.D. theses, journal articles, and books; multiple languages and continents; and decades.  What is nothing short of shocking is that these studies, almost all of them, from different research groups, including rancorously competing ones, have arrived at the same closely circumscribed set of grammars characterizing natural languages (Joshi, Vijay-Shanker, and Weir 1991; Vijay-Shanker and Weir 1994; Pullum and Gazdar 1982; Shieber 1985).

(Acquisition of natural language notably can't be tested "in the wild" except by observation. Natural language acquisition is notably unsupervised and rapid, when it is developmentally early; second language learning typically lacks the feeling of "effortlessness" that appears to be characteristic of developmental language acquisition. Researchers have constructed artificial sequences that variously conform to properties of natural languages in order to run more controlled experiments on acquisition. These "artificial grammar learning" (AGL) experiments have largely accorded with hypotheses about natural language learning (see, e.g., (Fitch and Friederici 2012) for review). Subjects are presented with "sentences" or "strings" that are generated by a set of rules corresponding to grammars of different types in the grammar hierarchy; these presented strings can lead to learning of regularities, which are often described as "implicit" learning – because the generative rules underlying the presented artificial "sentences" are never explicitly presented. Confusingly, in some experimental paradigms, the strings are just exposed to subjects, without training or feedback, whereas in other paradigms, subjects are told which strings are of one type and which are another (something that does not occur in natural language acquisition). The conflation of these highly different methodologies causes misunderstandings in the field. Nonetheless, certain types of evidence have been presented for very small artificial languages being treated in ways argued to be concordant with early-acquired full languages (Friederici, Steinhauer, and Pfeifer 2002).)

(In an unrelated controversy, it was recently argued that the native language of a particular tribe may have contained no embedded clauses. We here are solely concerned with human capacities. For studying the computational power of human cognition, the sole relevant point is that some tribe members were raised with Portuguese as their native language, and they did learn embedded clauses. These peoples' language capacities are the same as those of other humans. We are seeking the properties that all humans and only humans have, and the members of that tribe clearly have them.)

In simplest terms, so-called "context free" grammars are shown to be too small to explain some explicit examples from natural languages, whereas so-called "context sensitive" grammars are shown to be too broad, spuriously accepting utterances that are easily shown to not occur in natural languages. A grammar size that is formally greater than context free, and lesser than context sensitive, has been shown to fit the data. (In detail, analyses of some natural languages have concluded that context free grammars suffice (e.g., (Pullum and Gazdar 1982)); for some languages, context free is too small a class. No natural languages have ever been shown to be the full size of context sensitive grammars.)

*Thesis A.4) A core capacity?*
Counting, logic, perspective-taking; ... these brief references are in no way meant to constitute an exhaustive list; rather, they are meant as an introduction to a substantial category: well-studied non-human precursors to unique human abilities. As mentioned, many such abilities appear to follow this pattern:
  i)   an antecedent capability present in nonhumans and humans, and sometimes present in preliminary form during human ontogeny ("ancestral");
  ii)  a capability that is uniquely easy for humans to developmentally acquire, but very difficult or impossible for nonhumans ("all-and-only");
  iii) an advanced human capability requiring extensive training and external equipment ("adept").

These three stages or versions of an ability may successively depend on each other, building up via some form of added capacities that lift an organism from one level of capability to the next.  The intended focus is the category of 'all-and-only,' uniquely anthropic, capabilities.

In light of these seeming relations among capabilities, we forward the possibility of a core capacity, occurring solely in humans, surpassing nonhuman precursor abilities; that is universally present among humans, barring pathology.  That is: all humans, and only humans.  Such unique and ubiquitous abilities would be acquired along a developmental trajectory even without any explicit training, solely unsupervisedly, simply via exposure to the relevant data.

The next two sections first describe the unique all-and-only characteristics of human neuroanatomy, and then discuss and illustrate the automata hierarchy.  Then we will return to the central question:  where does human cognition reside in that hierarchy?

### B) Comparative neuroanatomy of humans and other primates
*Thesis B.1) Primate brains are allometrically uniform*
Human brains are in some ways evolutionary "kludges," amalgams of randomly appended faculties that have been accreted over roughly 200 million years of mammalian evolution, 4 million years of primate evolution, and then some hundreds of thousands of years of uniquely homo sapiens evolution (Lynch and Granger 2008; Striedter 2005).

However, when we address mammalian evolution in terms of ongoing adaptations, we face a stark puzzle: the range of structural variation differentiating mammal brains is extraordinarily limited; mammalian brains are so much alike, aside from simple size differences, that human brains are referred to in the literature as simply "scaled-up chimp brains" (Herculano-Houzel 2009; Herculano-Houzel et al. 2007).  The ongoing search for substantive differences between human and other brains keeps turning up seeming minutiae, with no apparent explanatory power over the extent of humans' difference from other primates (Rodriguez and Granger 2016).  That is, primate brain variation is miniscule, begging the question of where unique human abilities could be coming from.  As brain size scales across four orders of magnitude in the mammals, every component brain part stays intact, and, with phenomenal precision, even the exact ratios among brain parts (e.g., cortical to subcortical size ratios) vary in strictly predictable allometric correspondences.  The few specific deviations from this rule (e.g., olfactory cortex in primates) establish that evolution is quite capable of diverging from the rigid allometric constraints, and yet, save for these few exceptions, the allometry is stringently adhered to.

Thus, apart from these rare exceptions, brain organization has astoundingly little leeway to expand or contract some structures at the expense of others in response to presumed evolutionary pressures.  This represents a compelling constraint on interpretations of human brain specializations.

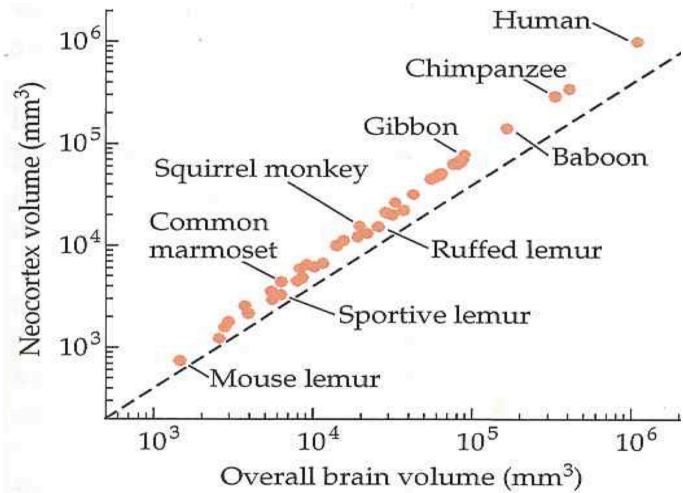

Figure 5. Neocortex (including posterior and anterior), with an allometric slope greater than 1, occupies an ever-larger percentage of the brain as overall primate brain size increases. (From Striedter '05).

Of crucial importance is the simple fact that evolutionary advances need not "optimize" anything, such as making mechanisms simpler or more efficient. Allometry startlingly illustrates the lack of substantive variation among brains and the almost unbelievably rigid consistency of size and connectivity relations throughout mammalian brain structures. (Indeed, it was not widely incorporated into the literature until the groundbreaking allometric publications of Finlay & Darlington (Finlay and Darlington 1995) and colleagues, building on the extensive metrics obtained by Stephan et al. (Stephan, Frahm, and Baron 1981; Stephan, Baron, and Frahm 1986).)

Some size, shape, or connectivity changes do occur beyond the allometrically-determined; such "mosaic" modifications are surprisingly rare, and, even when they occur, are very small (typically representing changes of less than a percent in predicted size).

*Thesis B.2) Human anterior cortex is almost an order of magnitude larger than for any other primate*
The allometric slope for neocortex is higher than one, that is, as brain size increases, neocortex grows to be a proportionately larger percentage of the brain. In general, the phylogenetically later the cortical region, the higher the slope (Finlay and Darlington 1995; Herculano-Houzel 2012). Anterior cortical regions such as dorsolateral prefrontal (dlPFC) and area 10 become proportionately even larger than other cortical areas, albeit in allometrically predictable fashion (Semendeferi et al. 2002).

For instance, the frontal pole, or Brodmann area 10, is argued to exist in all mammalian species (but see Preuss (Preuss 2000)). Area 10 is large in primates and is very disproportionately large in humans: it occupies about 0.74% of total brain volume in primates such as bonobos (and far smaller percentages in other mammals), but area 10 constitutes roughly 1.2% of human brains, all the while conforming exactly to the size that is predicted by simple known allometric relations among primate brains.

That is, simply by knowing the overall size of a human brain, we may allometrically predict the size of its components, such as area 10, with remarkable empirical precision (Fig 6) .

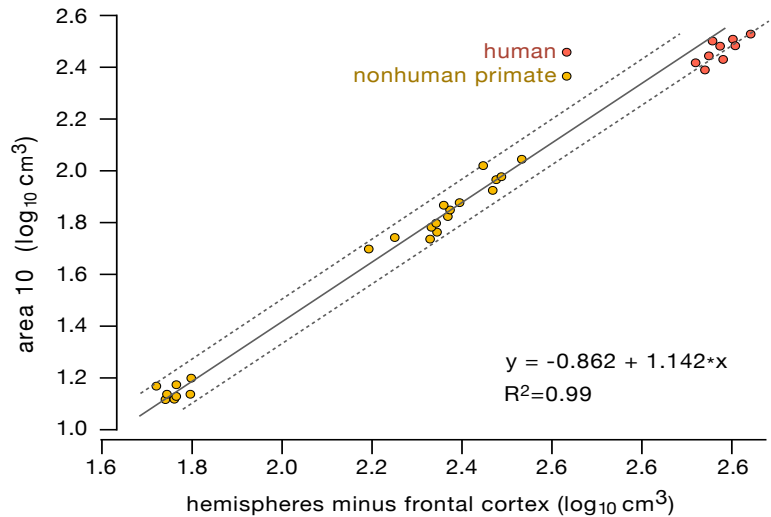

Figure 6. Logarithmic plot of area 10 cortex volume vs. remainder of telencephalon for six primates (gibbon, orangutan, gorilla, bonobo, chimp, human). The slope of greater than 1 is consistent across the primates; best-fit regression line is shown flanked by (dotted) prediction interval. The proprtion of the brain allocated to area 10 will substantially increase as overall brain size grows larger. (Adapted from Semendefari 02)

Some areas within a human brain are roughly the same size as those in the ape brain, whereas most are disproportionately much larger, due to the different allometric slopes for each brain region. The highest slopes belong to anterior cortical regions. The average brain area is roughly three times larger than that of an ape – but the anterior cortical regions such as dlPFC, area 10, and other prefrontal regions, are roughly eight times larger.

A leap of this size, almost an order of magnitude – specifically in brain areas that appear central to higher cognitive processes – is unprecedented among the primates. The changes emerge, as do almost all major changes among primates, from the standard rules of allometry; yet the sheer size of these anterior cortical regions makes them difficult to classify as gradual.

(Although Tattersall (Tattersall 2017) quite correctly points out that "brain size increase was a property of the Homo clade in general," he uses this to assert that, therefore, brain size increase "evidently had nothing to do with how modern Homo sapiens, specifically, did or does its cognitive business." By contrast, we are highlighting the fact that human brains represent a colossal leap of unprecedented size, which cannot be ruled out as having something very much to do with present-day humans' cognitive capabilities. The question of intermediate brains in our lineage still remains unanswered, but it cannot be used to discount the role of enormous allometric circuit expansion in our brains.)

Intermediate sizes did allometrically appear in other hominins, during the extraordinary burst of growth during the past 2.5 million years. Since those species are gone, we have no information about their possibly intermediate abilities, nor whether they too may have crossed a threshold that gave them language or other abilities that today are uniquely human. It is often noted that these abilities per se are insufficient to account for the radical success of humans; the abilities also had to occur within populations that were sufficiently large to enable communication to confer further advantages. Evidence indicates that other external technological milestones (such as the standard panoply of fire, cooking, agriculture) were also crucial steps – but of course none of them help explain the acquisition of new abilities, nor would any externalities have remotely sufficed in the absence of the uniquely human abilities that enabled them in the first place.

*Thesis B.3) Brain circuitry predominantly consists of cortical-subcortical loops*
Cortical-subcortical loop circuitry (including cortico-thalamic, cortico-striatal, and cortico-hippocampal loops) constitutes more than 98% of human brains by volume (and roughly 35% of rat brains). Figure 7 shows the repeated pattern of connectivity that is shared across all the primates (and all the mammals): cortex is invariably acting not alone, but in concert with multiple subcortical regions (Alexander and DeLong 1985).

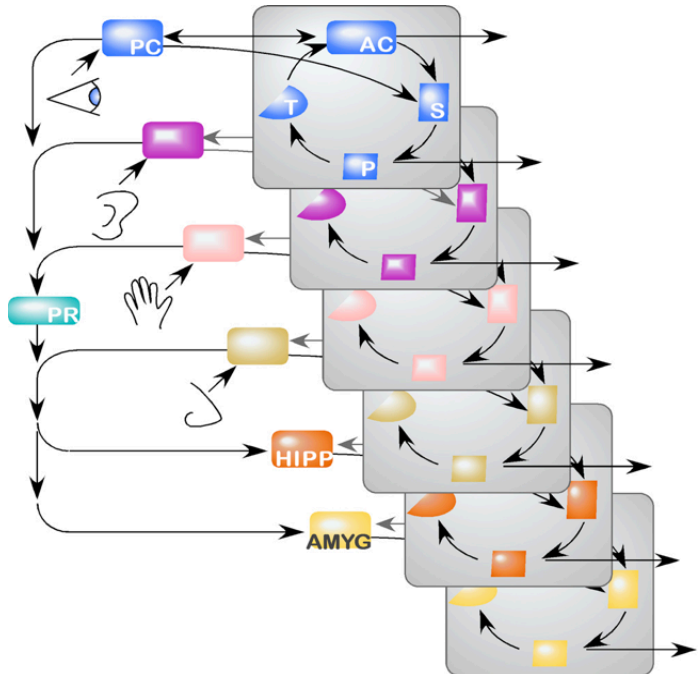

Figure 7. Strictly patterned cortical-subcortical loops dominate all forebrain circuitry. A given posterior cortical (PC) region innervates a corresponding anterior cortical (AC) region and a matched striatal (S) region; the AC area targets that striatal region, engaging a striatal-pallidal-thalamic (S,P,T) loop back to the same AC site; both pallidal and cortical (pyramidal tract) efferents activate lower motor structures (rightward arrows). Loops of this kind exist for visual, auditory, somatosensory, perirhinal, and limbic systems, establishing an extraordinarily widespread forebrain architecture, presumably organizing the vast majority of all cognitive function.

*Thesis B.4) Brain circuit simulations and analyses find cortico-thalamic loops compute grammars*

Extensive simulation and analysis of the physiological operation of these anatomical circuits concluded that individual loops created both categories (via topographic thalamocortical and cortico-cortical projections) and sequences (via the far more prevalent nontopographic or diffuse matrix thalamocortical and cortico-cortical connectivity (Herkenham 1986; Scheibel and Scheibel 1967; Barbas and Rempel-Clower 1997; Batardiere et al. 2002; Budd and Kisvarday 2012; Rockland 2004; Swadlow, Gusev, and Bezdudnaya 2002). The output of a given cortical area becomes input (both divergent and convergent) to other, downstream, regions, as well as receiving feedback from them. Producing categories and sequences, in cortico-cortical succession, yields sequences of categories, and categories of sequences of categories, etc. This hypothesized primary computation of thalamo-cortico-cortical circuitry (Swadlow, Gusev, and Bezdudnaya 2002; George and Hawkins 2009; Granger 2006; Rodriguez and Granger 2016) is formally equivalent to grammars.

Extended networks produce successively nested grammars of increasing depth (concordant with findings that increasingly long auditory patterns are selectively processed by successively downstream cortical regions). A single category at any one cortical locus can itself be part of another entire sequence of categories.

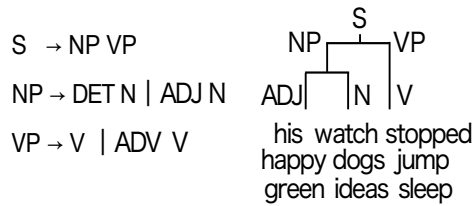

Figure 8. The category **S** expands to the sequence NP VP; these categories in turn expand to disjuncts of either DET N or ADJ N; or V or ADV V. Many combinatorially possible constructions can be composited; one such is shown at right, along with three sample renderings of the nonterminal items (S, NP, VP, N, etc) to terminals (in this instance, English words, but these can be arbitrary perceptual or cognitive objects (see text).

Figure 8 is a simple illustration of the hierarchical nature of nested grammars; these are illustrated with English word sequences, but extensive work has demonstrated how these structures occur in, and explain, similarly hierarchical processing of other stimuli from auditory and visual to abstract cognitive constructs (Zhu and Mumford 2006; Grenander 1996 ; Felzenszwalb 2011; Geman, Potter, and Chi 2002; Chandrashekar and Granger 2012).

In general, these formal grammars have no necessary relation to language; they are formal constructs arising from simulated cortical-subcortical performance, operating on any inputs that occur, beginning with simple sensory inputs and continuing downstream to construct ever-larger grammars.

A criticial characteristic of these systems is their production of a form of representation that is absent from typical artificial schemes whether unsupervised, supervised, or reinforcement based. That is the existence of "position-based categories." These are readily illustrated using linguistic examples, but as just described, their formulation is independent of language, and may apply equivalently to representation across other domains, from perception to reasoning. If "John hit a tree," and "John hit a ball," then "ball" and "tree" are in a category – not a similarity-based category, as in unsupervised systems, because ball and tree may have no shared or similar characteristics; not a labeled category, since no supervised information is presented; nor a reinforcement category, since no reinforcement information is presented. Rather, the category is based on its position, or role: it is something that can follow "hit," or something that can be hit (i.e., the "patient" role of hit). These "position-based" categories are of a different nature than the three others (similarity, labeled, reinforced). The ongoing development of sequence- and context-based learning systems suggest promising potential directions (Lund and Burgess 1996; Bullinaria and Levy 2007; Mikolov et al. 2013; Pennington, Socher, and Manning 2014).

Only some of these systems are generative, and as such they yield richer representations than simple "discriminative" mechanisms, yet generative systems typically are shown to be more computationally expensive than discriminative algorithms (Ng and Jordan 2002). It is notable that the grammar mechanisms derived from cortical-subcortical loop circuitry have been shown to exhibit the representational richness of generative mechanisms, yet are less computationally costly than corresponding discriminative algorithms (Chandrashekar and Granger 2012); this exceptional pairing of high computational efficacy with low computational cost is rare.

As already discussed, different formulations of grammars exhibit distinct tiers of computational power; evidence thus far indicates that the grammars described here, derived from cortical-subcortical circuitry, are all single-stack algorithms, which can range from simple visibly-pushdown automata up through higher-order indexed pushdown automata computing nested-stack grammars. The different forms apparently depend on the nature of cortico-hippocampal interactions, as described further below.

*Thesis B.5) Cortico-hippocampal interactions play a specific role in memory encoding*
Extensive work indicates that the hippocampus plays a critical role in the formation and retrieval of episodic memory (Eichenbaum 1998; Ekstrom and Ranganath 2018; Moscovitch et al. 2016; Wixted et al. 2018) incorporating information ranging across semantic, spatial, and temporal (often "what," "where," "when"); the mechanisms whereby such information is stored and organized, however, remain very poorly understood.

Encephalization (Jerison 1977) denotes the increase in relative size of phylogenetically later brain structures. Encephalization derives almost entirely from differences in allometric exponents: neocortical regions, and especially anterior neocortical regions, grow disproportionately larger than any other brain areas. As some areas grow large, such as frontal cortical areas, others grow proportionately smaller, such as most subcortical structures – very much including the hippocampus. In a small-brained mammal such as a hedgehog, the ratio of neocortex to hippocampus is just less than 2:1, i.e., the hippocampus is more than half the size of the entire neocortex. In a chimp, the hippocampus is 1/100 the size of the neocortex; in a human, it is 1/1000 the size of the neocortex.

No cortical area operates in the absence of tightly coupled cortical-subcortical loops (Figure 7). Cortico-hippocampal loops are of particular interest here. Hippocampal units have been found to produce remarkable response patterns when engaged in memory-dependent behavior such as learned navigation: neurons "replay" and "preplay" both forward and backward activation sequences that previously occurred during successive moves while navigating through a learned space (Foster and Wilson 2006; Nádasdy et al. 1999; Davidson, Kloosterman, and Wilson 2009), and, notably, in many other conditions that are not strictly navigation-based (Gupta et al. 2010; Logothetis et al. 2012; Colgin 2016).

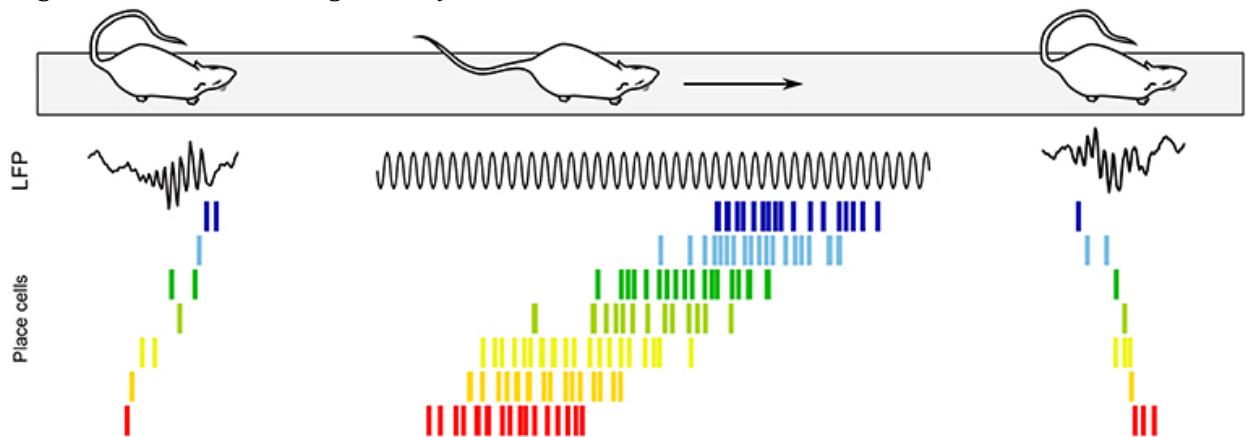

Figure 9. Place cell activity experienced during behavior (middle; 10s of seconds) are generated in highly compressed form (<100msec) during awake sharp-wave activity before (preplay) and after (replay; reverse sequence order) the behavior. Black traces show sample CA1 local field potentials (LFP).

The potential utility of the replay/preplay mechanism has been often conjectured in the literature; notably, this pattern of responses are consistent with what would be observed if sequential memory elements are being "pushed" onto and subsequently "popped" off of a stack: the elemental operations in the stack memories of automata (see section C for further stack discussion). These stack operations may enable the tracking of ordinal positions of a given item within a sequence. This could be of adaptive utility for navigating paths through space, for memory "indexing," and for multiple other sequence-dependent functions (O'Keefe and Nadel 1978; Eichenbaum 2000; Hafting et al. 2005; Papp, Witter, and Treves 2007).

The time spans for neural activity are orders of magnitude different than those for behavior. The former take tens or hundreds of milliseconds; the latter can extend across seconds or minutes. The unique long-lasting activity of the highly unusual recurrent collateral anatomy of hippocampal field CA3 may serve to "gear" these distinct time spans, enabling neural persistence over behavior-length durations (Cox et al. 2019).

The unique properties of hippocampal machinery apparently are critical to episodic memory storage and retrieval; the hippocampus is capable of some operation that is evidently absent in neocortex, causing the hippocampus to be a bottleneck in episodic memory processing. As brains grow larger and the hippocampus becomes a proportionally far smaller component, that bottleneck becomes more restrictive. This may indicate an important distinction between normal small-brained mammals versus those with extremely large brains and their attendant proportion changes. Some other instances of encephalization have the effect of rendering phylogenetically older regions largely superfluous: lesioning the superior colliculus in a rat severely impairs its visual abilities, whereas corresponding lesions in primates result in highly subtle impairments at most (Lawler and Cowey 1986). In the case of the hippocampus, however, although some hippocampal functions are arguably supplanted by cortical enlargement, nonetheless other functions evidently remain dependent on an intact hippocampus.

The two functions that will be highlighted here (in the section below sketching the dependence of episodic memory on hippocampus) are timing and stack storage. These roles of the hippocampus enable the encoding of extended temporal information, likely because its stack-like operations (discussed below) are somehow useful in those encodings. In order to continue to organize time sequence information, the hippocampus cannot be eliminated from the critical path of memory storage even when it becomes so small that it is a severe bottleneck in memory processing.

Overall, we conjecture that human cortex grows so large that its cortico-hippocampal operations not only supplement, but largely supplant, the form of memory storage and retrieval that dominate in smaller brained mammals. Cortical operations cannot fully supersede hippocampal operations due to a continued dependence on unique operations – simple stack operations – that are performed by hippocampus; cortex is able to perform the equivalent of nesting the resulting stacks, but cannot produce the stacks themselves without cortico-hippocampal interaction. This combination of circumstances (overlarge cortex, still-required hippocampal operation) enables humans to generate comparatively enormous cortically based memory capabilities (which we posit underlies the expansion to nested-stack operation), while retaining dependence on the hippocampus.

*Thesis B.6) Seemingly saltatory evolutionary jumps are consistent with allometry*

What is the nature of "jumps" from pre-human abilities to human abilities?  The seemingly saltatory leap between our language abilities versus any preceding abilities dichotomizes research: either language grew gradually, with intermediate forms, over long time periods, or it leaped into being "catastrophically" (Bickerton 1995) with the appearance of some relatively compact mutation that abruptly enabled language to emerge (Hauser et al. 2014; Tattersall 2017; Boeckx and Benitez-Burraco 2014; Suddendorf 2013).

How many before us in the genus Homo had neocortex-to-hippocampus ratios of substantially greater than 95%, i.e., neocortical sizes that almost entirely overwhelmed the hippocampus?  In such thoroughly encephalized species, we posit that cortical expropriation of what had been hippocampal operations, generated the first nested stacks.  Allometrically estimated brain structure sizes in supposed H. erectus and H. heidelbergensis yield intermediate neocortex/hippocampus ratios.  It is not until these structures become the default memory mechanisms that previously single unrestricted-stack automata (which generate context free grammars) became secondary to cortical nested stack systems, surpassing context-free grammars.

The *Homo* fossil record contains either punctate jumps or exceptionally fast "gradual" bursts of growth (Lynch and Granger 2008).  All in this ancestry are either extinct or absorbed into us; it is unknown whether any in the lineage exhibited our supposed "all-and-only" traits, or precursors of them.  Scant archaeological evidence has been used variously to strongly distinguish prehuman abilities from our own (Tattersall 2017) or to identify possible continuities among them (McBrearty and Brooks 1999).  Some evidence suggests extremely small and localized early populations in the Homo line.  It is crucial not to conflate the presence of intelligent creatures, on one hand, with possible cultural effects of many such creatures, on the other (McBrearty and Brooks 1999) (Nakahashi 2013).

## C)  Nature and application of automata and formal grammars

*Thesis C.1) The essential differences among automata are their memory systems*

All automata are simply FSMs with different types of memories.  Turing machines included.  So the essential question devolves to: what type of memory system?

As mentioned, the naming scheme for these automata and grammars is far from intuitive.  The sequence of automata begins with finite state machine, proceeds through restricted- and then unrestricted-stack machines, to nested-stack machines, bounded- and then unbounded-stack machines, and then Turing-complete machines.  It of course is non-obvious from the names alone that a "bounded" machine is more powerful than a "nested" machine, which in turn is more powerful than an "unrestricted" single stack machine. The terminology arises from the literature in formal automata theory, and does not correspond readily with nomenclature from other fields.

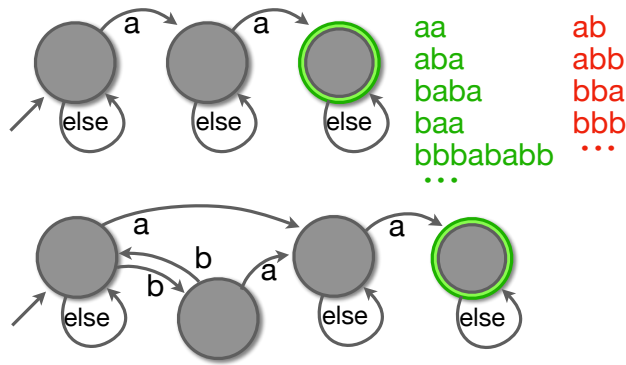

Figure 10. Finite state machines and equivalence classes. (Top) Initial input stimulus a or b arrives (at left) to the input state α. Next input arrives; if it is an "a," the machine transitions to state $\beta$; if it is not an "a" then no state transition occurs ("else"). Subsequent arrival of an "a" transitions the FSM to state γ, which is a halting state (double circle). Subsequent inputs have no further effect ("else"). (Right) Input sequences "accepted" (green) by the FSM, i.e., arriving at a halting state, or "rejected" (red; not arriving at a halting state). (Bottom) Another FSM, with different states and transitions, accepts and rejects the same sequences as the top FSM. Although different, the two are equivalent automata.

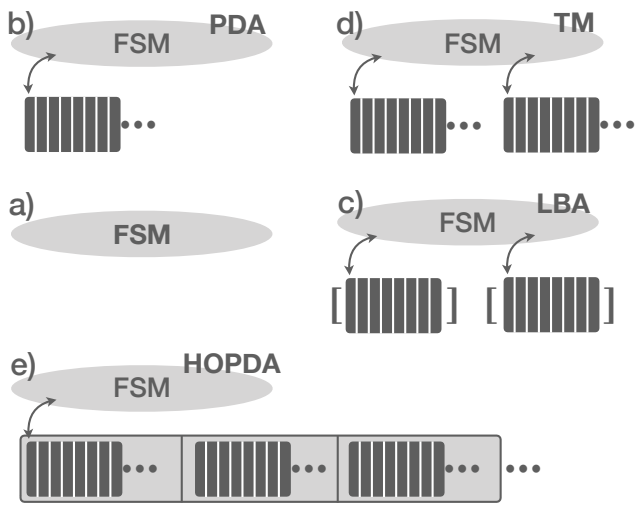

Figure 11. The architecture of automata. All automata contain a finite state machine (FSM; see Figure FINITESTATEfig) and some form of stack memory. a) An FSM alone without memory produces Regular grammars. b) Adding a **single** unrestricted stack produces a pushdown automaton (PDA), which computes context-free (CF) grammars. c,d) Adding **multiple** independently-accessible stacks either produces either a linear bounded automaton (LBA) if the stacks are restricted in length, computing context-sensitive (CS) grammars, or a Turing machine (TM) if the stacks are unrestricted, computing recursively enumerable (RE) grammars. e) Higher-order PDAs (HOPDAs) incorporate stacks of stacks (nested stacks) to produce larger grammars than context free, but less than context sensitive. These are the most powerful form of **single** stack automata.

Similarly, the names of the formal grammars that correspond to these machines may seem almost misleading. Recursively enumerable grammars (computed by a Turing machine) are a strictly larger class than so-called "context-sensitive" grammars (computed by bounded-stack automata), which in turn are strictly larger than indexed, combinatory, and tree-adjoining grammars (computed by nested-stack automata or HOPDAs); these in turn all are larger than "context-free" grammars (computed by counterintuitively-named "unrestricted-stack" automata); these finally are larger than visibly-pushdown grammars (restricted-stack automata) and regular grammars (computed by simple finite state automata with no stacks).

Colloquially, we may think that, since context clearly matters in language, and language is thus indeed context-sensitive, that this is a feature that is required in the formal characterization of natural language; but the formal system termed "context sensitive" is just a (perhaps not ideally chosen) descriptor for a mathematical entity with specific characteristics in the automata hierarchy.  As mentioned, these context-sensitive formal systems produce linguistic constructs that have been shown not to occur in human natural language: put another way, context-sensitive grammars are too powerful to be natural language machines.  Actual human natural languages are less powerful than context-sensitive formal grammars.  And context-sensitive grammars are less powerful than "recursively enumerable" grammars, produced by Turing machines.  (One of the compelling side effects of the restriction on natural language grammars is that there are "impossible" languages, i.e., languages that we can formally conceive of but that do not and, arguably, cannot occur as an empirical human tongue (Moro 2008).

Notably, if we begin with, say, a so-called "unrestricted stack" automaton, which recognizes context-free grammars, and we add to it a system of nested stacks, doing so is equivalent to transforming or upgrading the original system from an unrestricted-stack system to a nested-stack system, which is a system of strictly greater computational power.

Formal stacks are defined as last-in first-out, i.e., the entity most recently stored is the first retrieved.  (If we store (or "push") the sequence a,b,c onto the stack, then we retrieve ("pop") them back off of the stack in the order c,b,a; by contrast, standard "queues" are first-in first-out, i.e., retrieval is in the order of storage (push a,b,c; pop a,b,c).)  Formal stack behavior is not intended to be directly commensurate with cortico-hippocampal operation, yet it is noteworthy that measurement of hippocampal units are reported to yield both forward and backward "replay" and "preplay" constituting rapid sequences of activations that previously occurred (and presumably were stored) during behaviors, such as successive locations arrived at in an environment.  For instance, if locations a,b,c occurred in sequence during navigation, then a,b,c and c,b,a may be seen to rapidly occur during hippocampal replay (Nádasdy et al. 1999; Foster and Wilson 2006; Davidson, Kloosterman, and Wilson 2009; Gupta et al. 2010).  It has been noted that these memory behaviors can be useful not only in terms of navigation, but also in terms of general memory "indexing," and other potential sequence-dependent memory operations.

As discussed earlier, many different formulations can carry out the same computations.  Both an electronic adding machine and a mechanical adding machine can add, even though they may share no common components or operational steps.  Analyses aimed at identifying computational power (such as Turing equivalence) must always be viewed in terms of such equivalence classes.  Further distinctions require other analyses, beyond those of just the automata hierarchy.

It is crucial to note that any task that is performable by a Turing machine, the top of the automata hierarchy, can be perfectly carried out by an appropriately configured finite state machine (FSM), the bottom of that hierarchy.  In what possible sense, then, could these be considered to be machines with different computational powers?  The answer is not a minor detail, but rather is the essence of the automata hierarchy: any change to the task can still be carried out by the Turing machine, whereas even slight changes may cause the FSM to fail.  The FSM can solely do tasks for which it is rigidly configured; it has literally no memory beyond its current state.  FSMs are, intuitively, all hard-coded information; all strictly innate, pre-specified rules.  Those rules may have substantial generality and power: a single FSM could be built to play chess or Go.  What a Turing machine has, by contrast, is a memory from which it can store and retrieve information.  It can thus be reconfigured on the fly, to perform differently based on what it has seen and stored.  Thus, to tell apart an FSM from a Turing machine (or any of the intermediate automata between these extremes in the hierarchy), requires careful testing and extrapolation, as further discussed below.

*Thesis C.2) Data indicate at most context-free capabilities for all nonhuman animals*
Given this formal hierarchy, what can it say about any specific empirical behaviors?  Researchers study corpora of data such as recorded human spoken language, or bird calls, and attempt to identify a fit between data and model.

Crucially, no amount of experimental evidence can ever entirely confirm a hypothesis about the fit between an automaton and a body of data.  As mentioned, any task can be simulated perfectly by a finite state machine with no stack memory whatsoever.  On one hand, that finite state machine may be enormous, and on the other hand, the very next dollop of data may fail to fit it: a classic instance of overfitting, as can occur with any model.  Instead, studies attempt to narrow in on a model from two directions: i) identifying classes of data in the corpus that provide evidence that a particular model is insufficient, and ii) generating model data to collect evidence that the model is more powerful than it needs to be to account for the corpus.

Thus, for instance, the utterances that would be produced by a context-sensitive grammar can be simulated by a sufficiently large context-free grammar model.  This risk of overfitting leads to a necessary conservatism in the evaluation of claims for particular models (Chomsky 1975).  Examinations of utterances have routinely proceeded in light of this admonition; the resulting literature nonetheless exhibits a consistency of findings that is persuasive.

Studies of corpora of nonhuman communication (calls of primates, avians, cetaceans, etc) repeatedly find them to be context-free or simpler.  There are analogic references to human language in some studies (Abe and Watanabe 2011; Beckers et al. 2012), but the analogies are difficult to substantiate.  Studies that carefully analyze bird song, identify it with finite state machines (FSMs) (Gentner and Hulse 1998; Berwick et al. 2011), or even sub-FSM systems (Berwick and Pilato 1987; Beckers et al. 2012).  (Sometimes the terminology "compositional syntax" is invoked; this nomenclature is indeterminate with respect to the grammar hierarchy.  Compositionality is exhibited by Turing machines --- but also by the regular grammars of simple finite state machines, and any machines in between.  The key is to identify the detailed nature of what is purportedly being composed.)

*Thesis C.3) Data overwhelmingly indicate nested stack automata for human language syntax*

Extensive studies of human language capacity have been carried out over the course of decades, on multiple natural languages across all continents.  The investigations repeatedly arrive at a fixed set of automata variants: tree-adjoining, head grammars, combinatory categorial grammars, and linear indexed grammars (Pollard 1984; Joshi and Schabes 1997; Shieber 1985; Joshi, Vijay-Shanker, and Weir 1991; Steedman 1987; Baldridge 2002; Vijay-Shanker and Weir 1994).  This seemingly disparate range of findings are in fact stunningly circumscribed: all are more powerful than context-free grammars, and less powerful than context-sensitive grammars.  (Some have pointed out that the majority of natural language practice is context free, though they acknowledge that supra-context-free usages demonstrably exist; see (Pullum and Gazdar 1982).  These characterizations all inhabit the portion of the automata hierarchy corresponding to "nested stack" automata, producing the general class of "indexed" grammars (Aho 1968, 1969; damm 1982).  That is, all of the grammars identified from empirical studies of human natural languages arrive at special cases of this distinct category.

These findings appear extraordinarily unlikely to be simply coincidental; they appear to be compelling evidence that the mechanisms underlying human language exhibit this specific narrow range of computational capacity, whereas analyses of the corresponding mechanisms underlying other animals' communication systems indicate lesser computational power. (No in-depth review of these extensive technical findings is offered here; references such as (Vijay-Shanker and Weir 1994; Joshi, Vijay-Shanker, and Weir 1991) provide elaborate accounts of these findings.)

*Thesis C.4) Nested-stack automata are a "local ceiling" in the automata hierarchy*

Quite independent of any empirical considerations whatsoever (such as natural language), there are reasons to posit that nested stack automata constitute a fixed limit constraining the natural growth of grammars. Specifically, nested stack automata are the most powerful systems that can be mathematically achieved using strictly single (serial) stacks; all more-powerful automata (context sensitive and up) formally require simultaneous parallel access to multiple independent stacks. This represents a highly persuasive natural break point in automata growth.

Many standard depictions of the automata hierarchy contain Type 0 to Type 3 automata, which respectively produce regular grammars (Type 3), context free grammars (Type 2), context sensitive grammars (Type 1), and recursively enumerable grammars (Type 0). Since the introduction of those tiers (Chomsky 1975), there have been multiple elaborations of the hierarchy. Two such elaborations constitute fundamental extensions: i) the realm of "subregular" grammars, specifying distinctions among the smallest and simplest memory-free automata (McNaughton and Papert 1971; Heinz), and ii) the discovery and formal characterization of grammars that are at a strict intermediate level between context-free and context-sensitive, as detailed above. The subregular grammars are neglected in the current work, not out of lack of interest, but because they are beyond the scope of the (already lengthy) distinctions targeted herein. The intermediate supra-context-free and sub-context-sensitive grammars are a primary focus of the work presented here.

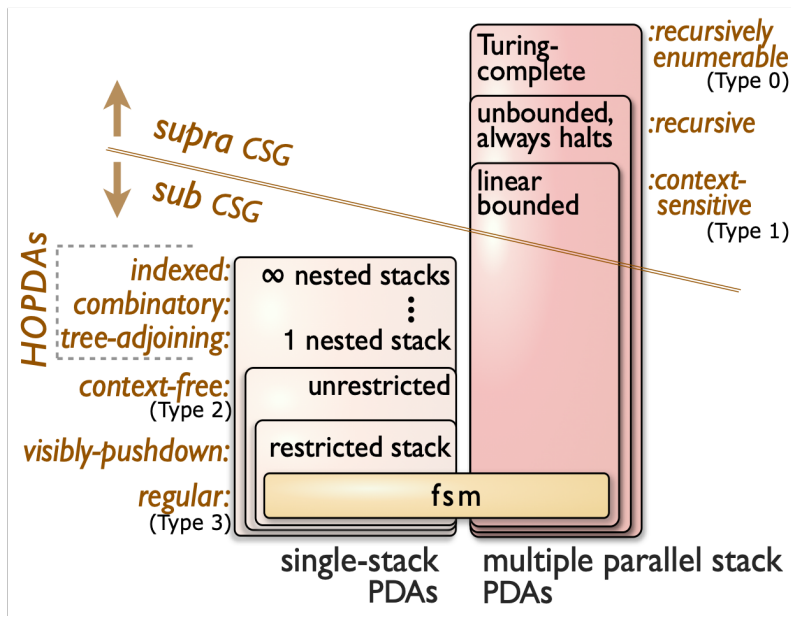

Figure 12. Expanded view of the automata and grammar hierarchies (compare Figure 3, which lacks the distinction between single versus multiple pushdown stacks). The emphasis here is on the divide between mechanisms containing only **single** pushdown stacks from those with **multiple independent** pushdown stacks: it is this essential change to the mechanisms of automata memory that fundamentally separates sub-context-sensitive grammars (CSG) from supra-CSG grammars.

The HOPDA automata and their attendant nested grammars were initially identified by Aho (Aho 1968) and elaborated by many others (Hopcroft and Ullman 1969; Maslov 1974; Joshi, Levy, and Takahashi 1975). In brief, initial work illustrated naturally occurring human language exceptions to context free practice, suggesting that context free grammars were insufficient to fully characterize language. The development of further formalisms such as tree-adjoining, combinatory, and head grammars, endeavored to specify mechanisms that could account for the empirical data in human language. The full characterization of nested stack automata become enriched over time, clarifying that these did indeed constitute a natural class of automata that are stronger than context free and lesser than context sensitive.

**IV. Ongoing issues**

*a) Innate means the capability to convert experience into particular abilities*
All humans, and only humans, acquire certain abilities, such as human counting, human language, human mental time travel, and others.

Nonhuman animals acquire lesser abilities that may be precursors to these human abilities – but animals clearly do not acquire this human constellation of abilities.

Conversely, no humans effortlessly and ubiquitously learn writing, or logic, or mathematics. We may almost effortlessly acquire the pre-human forms of these (such as numerosity), but we do not become capable of human-level writing or math unless we are provided with extensive external resources, memory extensions (chalk, pencil, paper, textbooks), and arduous training. (This distinction perhaps may be reminiscent of what is needed for other animals to learn even paltry subsets of human language).

Something in our brains is there to enable this. Whatever form it takes, the twin questions are:
   i)   Why does this system enable *effortless* language learning, but still requires hugely *effortful* learning of so many other tasks, such as writing, logic, math?
   ii)  Why is this system *absent* in all other animals; how does it appear in just us?

The term "innate" can draw audible gasps from certain conference audiences (and reviewers). There is a need to refer to precisely those capabilities that are unique, ubiquitous, and effortless to us; i.e., that are acquired by all and only humans. What is it in us that enables us to hear natural language and acquire it (by whatever means) whereas other animals do not do so?

The term "innate" in this usage certainly does not mean "pre-built-in"; it means that there is something in every human, before they have any linguistic experience, that enables them to take such experience when it arrives, and to effortlessly acquire it with its complex rules -- despite a) being unable to correspondingly acquire other similarly-complex systems, such as formal logic, without effort, and b) despite all other animals being unable to take those same inputs, under the same circumstances, and acquire the same linguistic system. Something is specifically different in humans.

It has been argued to be everything from a patchwork quilt of built-in templates (somehow evolutionarily acquired in very recent evolutionary history, while showing as yet no explanatory indicator of those differences in any anatomy, chemistry, or genetics), to a barely-defined capability to construct hierarchies (Yang et al. 2017), without any careful characterizations distinguishing the linguistic versions of those hierarchies from the many other hierarchies that humans (and even other animals) copiously exhibit (Jackendoff 2011).

*b) Some intuitions about antecedent, intrinsically human, and expert capabilities*
Humans with external memory are qualitatively different from humans alone

We are far from the first to posit that human brains can be formally treated as complex computational systems (Turing 1936). How powerful are human brains? Humans learn and remember, and can be said to read from and write to an intrinsic memory store. Do they have the computational power of full Turing machines, at the top of the automata hierarchy?

A human, of course, invented Turing machines, and we can build Turing-machine-equivalent computers; mustn't we therefore be at least as powerful as the things we build?

Any questions about specific machines, of course, depend on the formal definitions of those machines. In the automata hierarchy, the defining distinctions among differing systems crucially arise from different memory storage and retrieval mechanisms.

In the deceptively simple card game of "concentration," i) cards are placed face down in a spatial array; ii) randomly selected cards are transiently viewed to expose rank (e.g., Q) and color (red/black); and iii) matching pairs, at any locations, are removed, with the aim of eventually removing all cards. The features of previously-seen hidden cards must be remembered at their locations, and a card that was already removed is no longer eligible to match.

If arbitrary unlimited memory is available, the game is a trivial table lookup. Yet humans are challenged by the game, making errors both of omission (missing a potential match) and commission (erroneously committing to a hypothesized match that turns out false). Are humans Turing machines that somehow lack the ability to adeptly corral their infinite tape memories sufficiently well to remember several cards that they have recently seen? It is notable that, simply equipped with a paper and pencil, any human can effortlessly play the game with no errors, eliminating the difficulties imposed by memory limitations. Could a chimp learn to play the game by utilizing an external memory of this kind? The question is at present untested.

Similarly, a typical schoolchild can multiply two ten-digit numbers with pencil and paper. Carrying out the same operation in one's head is beyond the capacity of most students. Or professors.

Myriad such examples illustrate the stark differences between humans with only their own intrinsic abilities, versus humans with additional, carefully trained, external mechanisms. These simple illustrative examples on their own do not make the case; the case is made by the carefully assembled arguments constructed in this paper.

*c) There may be specialized circuitry for all-and-only human abilities – and there may not be.*
Hof and colleagues (Sherwood et al. 2012) warn that although "it is to be expected that some modifications to neocortical organization will be consequences of overall brain-size expansion ... other features, however, may prove to have emerged independent of encephalization." They further say that "to some extent, explaining human cognitive uniqueness as merely a by-product of encephalization, absolute brain size, or total numbers of neurons reflects the infancy of studies in evolutionary neuroscience." A plethora of studies attempt to find unique human characteristics that are not simply due to allometry or other scaling issues. As mentioned earlier, seemingly unique human specializations, such as Von Economo or roseship neurons, or ASPM or FOXP2 alleles, have repeatedly been found to occur in nonhumans – and even if they were uniquely human, they would be but a small step on the way to actual explanatory accounts of how their presence somehow generates entirely new mechanisms to give all-and-only abilities to humans.

The search for punctate biological differences may eventually succeed. But it may turn out that the differences are distributed and cumulative – far from punctate. And it may even turn out, as Hof and many others have noted, that there is something that arises during brain-size expansion that is not just random quantitative increase, but differential increases in highly specific circuitries, that generates fundamental changes in the modes of operation of the brain, mechanistically conferring new mental capacities.

It is unknown whether this is the moment that we should be looking to such devices as explanatory mechanisms. Perhaps the search for punctate differences, though exhausting, is not exhausted. A remarkable number of constraints have been identified, and a trove of previously-forwarded hypotheses have been all but ruled out, but in some ways, these are indeed still early days for brain evolution.

We do not turn to scaling issues out of any despair at the ongoing and necessary search for new brain characteristics. Rather, we affirmatively posit that human unique and ubiquitous abilities, very much including language, arise as a (huge and crucial) qualitative difference originating from a (colossal) quantitative change. As we have seen, when these abilities are quantified, as in human language, they cry out specifically for the scaling of memory mechanisms as their fundamental explanation.

Where there are human and prehuman versions of an ability, the relations between these versions will continue to be of interest. Whatever the terminology, it is crucial to make appropriate distinctions. To aggregate effortless and effortful capabilities all together is simply to fail to recognize a distinction. If the argument is that the differences are minor, that case can only be made once the distinctions are clearly articulated. We here emphasize that there are profound distinctions: between human capacities that come extraordinarily naturally for us and for no others (such as natural language and counting); abilities that come naturally both to us and to other animals (e.g., numerosity); and abilities that do not come naturally even for the vast majority of humans (writing, formal logic). Our aim is to distinguish among these clearly-separable capacities, and to attempt to characterize their differences.

The conclusions established here arise in large part from the well documented, but nonetheless extremely surprising, evolutionary adherence to allometry: the constituent components of a mammalian brain are strikingly and accurately predictable via straightforward allometric equations, as opposed to any evidence of any evolutionary pressures. There is no evidence of substantial contributions needed from natural selection (Finlay and Darlington 1995; Herculano-Houzel 2009, 2012; Charvet, Striedter, and Finlay 2011).

Hauser et al. (Hauser et al. 2014) argue that size alone is not sufficient for human language, citing two key facts:
- autistics with larger brains often nonetheless have language deficits;
- children with one hemisphere removed prior to full acquisition of language often display normal language expression and comprehension.

The former point simply shows that size is not sufficient, though it may be necessary; the latter purports to argue that size is not necessary. Autistics may simply have additional mechanisms in place that actively interfere with otherwise-intact language systems; these and other pathologies are miswirings, which of course can prevent language (and other human abilities) over and above the proper sizes. And children who have undergone certain surgical procedures are nonetheless encephalized to a greater extent than any other primate, and still exhibit an enormous cortical-subcortical ratio. That is our predictor of sufficient circuitry. Not "size alone." Language (and possibly other all-and-only human abilities) are "presumably the product of a complex and specific internal wiring, and not simply some slowly-evolved gross byproduct of increasing encephalization." It is unclear why the meticulous specificity of allometric encephalization may be dismissed as "gross," as though to make it appear that it is insufficiently exacting to accomplish the change from one kind of automaton to another. Specific brain circuitry, with distinct circuits occurring in particular ratios, "gross" or not, may constitute the necessary conditions for the emergence of human language.

Mammalian thalamocortical loops compute grammars, but in small brains, these have correspondingly small, hippocampally-limited memory storage stack mechanisms, limiting them to sub-context-free grammars. All mammalian brains, including humans, are dependent on hippocampal stacks in cortico-hippocampal loops, but the huge cortical-subcortical ratio in humans enables supplementation of hippocampal stacks via cortical storage. This changes the nature of our working memory capacity (Brady, Störmer, and Alvarez 2016), enabling encephalized cortical performance of otherwise hippocampus-only memory storage. We have here hypothesized that the result is the hierarchical cortical ability to nest stacks, which the hippocampal memory system cannot do. Nested-stack augmentation of the smaller-brained context-free ability enables the computational power of nested stack automata, which in various analyses have corresponded to "mildly context-sensitive," "tree-adjoining," and "categorial combinatory" grammars: all stronger than context free and weaker than context sensitive.

It is this specific circuitry, in these altered proportions – not just indiscriminate increased size of any old unspecified circuits – that rises to the level of nested stack automata. It is this change that confers human language abilities as well as other all-and-only human abilities – those capacities possessed by all humans, and only humans.

*d) Cortico-hippocampal loops compute the equivalent of nested stacks, critical for episodic memory*
In a recent experiment (Wixted et al. 2018) individuals took a 20-30 minute walk across an unfamiliar university campus during which several events occurred. Recall of the individual objects that were seen, their locations, and the order in which they occurred was assessed directly afterwards. Subjects with hippocampal damage were still somewhat able to report what they saw, and where they saw it, but they had no ability to describe the sequence in which they encountered these objects along their walking path.

This ability to encode time and sequence information (I saw the building before I saw the large tree; and after those, I saw the truck) apparently must rely on mechanisms within the specialized circuitry of the several hippocampal fields, that evidently are absent from neocortex. Intact cortico-hippocampal interactions are, for some reason, required to deal with episodic information.

In light of the extreme hippocampal bottleneck in humans, the findings of Alvarez et al. (Brady, Störmer, and Alvarez 2016) may lead to further predictions with respect to time-varying data. They found that working memory capacity was dependent on stored real-world knowledge of the objects to be held in memory, possibly suggesting cortical (real-world memory) schemas created on the fly that accumulate and hold information over longer time periods than would be possible using solely hippocampal-dependent mechanisms.  We suggest that such findings will not hold for moving stimuli, e.g., movies of objects occurring in spatial locations in a virtual environment: organization of these stimuli ought to continue to be dependent on the bottleneck of hippocampal mechanisms, so tests of sequential order retained in working memory should show a lower capacity than experiments, like those in Alvarez et al., that do not test for retained sequential order information.

We have emphasized that all automata differences arise solely from differences in the precise form of their memory systems.  Even Turing machines, at the top of the automata hierarchy, are just FSMs (bottom of the hierarchy), plus particular forms of memory manipulation.  Importantly, as mentioned, vast numbers of different physical systems can equate to the same class of automaton; the abstract mechanisms (FSM+memory) capture not the apparatus but the equivalence class (as in the simple example of Figure 10).

We may articulate the "cortical subcortical nested stack supplementation" (CSNS) hypothesis:
> *Human cortices grow so disproportionately large that they supplement hippocampal*
> *stack memories with cortical memory structures, producing nested-stack grammars.*

The hippocampi are still a gateway, perhaps because they have unique mechanisms for gearing time durations between brief (millisecond) neurophysiological time and extended (seconds or longer) behavioral time.  They are still on the critical path, but now constitute an ineluctable hippocampal bottleneck.

*e) On "scala naturae" fallacies*
If ants or bees can perform advanced navigation, then they might seem already to be high in the grammar hierarchy – and in general, some specific behavioral accomplishment may seem to be associated with some particular automaton.  This surrogate "*scala naturae*" can be seductively misleading.  The fallacy is that any given complex grammar can be simulated perfectly by a weaker grammar, even by a (sufficiently large) finite state machine alone – up to some fixed embedding depth.

Put simply, all of the states that would arise from memory retrieval in a higher-level automaton can simply be pre-created and made into explicit extra states in a state machine, with or without memory.  Thus, for any single set of tasks given in advance, a finite state machine can be fashioned to perform the tasks, no matter how computationally demanding.  Any task that, in full, would require a Turing machine, can be simulated in part, up to some pre-established precision.  The moment a new task is introduced, outside the set of previous fixed tasks, then the Turing machine can nonetheless perform it whereas the FSM will fail.

It is far from trivial to discover whether some given operation (e.g., a critical syntactic operation) may require the full power of a particular automaton, or only require a proper subset of such machines.  Ants may well be navigating via large FSMs, exhibiting prowess on tasks that are "built in," and abruptly failing in tasks that would otherwise seem to be equivalently demanding.

As has long been noted (Chomsky 1975), experimental evidence may at best weaken a hypothesis, but is insufficient to fully verify (let alone discover) the existence of particular computational power.  This uncertainty of identifying the exact power required for a task, and the corresponding potential pitfall of an imagined "scala naturae," warn of possible spurious interpretations.  These are of course nonetheless not to be taken as arguments against the simple reality of the extensively studied automata hierarchy, nor arguments against powers gained through evolutionary advance.

*f) Equivalence classes are not simply conflating abstract formalisms directly with brain mechanisms*
The levels of implementation, algorithm, and computation are of course separable. Mathematical formalisms of stacks might have turned out to be equivalently implemented via entirely different mechanisms looking nothing like mathematical stacks. It is somewhat surprising that in this case, based on the hippocampal preplay/replay literature, the biological version and mathematical version may turn out to closely correspond. Certainly nothing rules this out, and the evidence arises from biological studies of the hippocampus, not from stacks being retrofitted or arbitrarily assigned to the hippocampus. (See also (Gallistel and King 2011) for further discussion of the relationships between formal mechanisms and potential neural implementations).

The allometric evidence presented here is an instance of sheer bottom-up analytic treatment of detailed cortical-subcortical circuitry, which quite unexpectedly led to identification of grammars and grammar growth – not at all the other way around.

The work did not start with hierarchies, nor with stacks, nor with language at all. These findings are in no way artificially superimposing abstractions onto neuroscience; they arise from bottom-up simulation and analytic studies of telencephalic circuitry, that turn out to strongly implicate formal grammars as emergent circuit mechanisms. Those formal grammars were in no way related to language, but rather to elemental brain operations, and it was quite unforeseen, and unlooked-for, that the findings then turned out to be concordant with linguistic phenomena (in addition to a range of nonlinguistic neurobiological, behavioral, and evolutionary data).

*g) A vast set of candidate anatomies for human language are consistent with pure allometric growth*
Several researchers (see, e.g., (Berwick and Chomsky 2016; Friederici et al. 2017)) have pointed out that growth of arcuate fasciculus and superior longitudinal fasciculus connects BA44 and 45, temporal cortical regions, and supramarginal cortex, completing an anatomical loop that is unique to humans. This could have been argued to be an adaptation. But the size of the arcuate fasciculus, like other structures and pathways, are directly allometrically determinable from overall brain size. Specifically, the human arcuate fasciculus is consistent with the size expected for a primate with a brain of human size. The connectivity of these axon tracts thus only arises in brains of a given size. This is an additional example of specific circuitry in combination with sufficient allometric growth. (Moreover, the arcuate fasciculus alone is far from the only anatomical story in language capacity; see, e.g., (Hagoort 2019) for recent review).

Beyond these questions of comparative anatomy are, of course, questions of mechanism. Computational analyses show *how* allometric growth can yield ostensibly saltatory changes, as increased stack mechanisms (already in place, but now allometrically enlarged) incrementally advance through the grammar hierarchy (Rodriguez and Granger 2016). By contrast, if a theory requires that a novel computation, not previously present in the brain, now arises in some specific brain area(s) (BA44, or others in the arcuate loop), such theory requires far more compelling evidence and mechanistic analyses before it moves from being a statistical observation to being a potential explanatory hypothesis.

*h) Neither simply gradualist nor simply catastrophist*
Tattersall (Tattersall 2017) dichotomizes the possible rise of human language: either it "came into existence gradually, over a vast period of time, as natural selection steadily exerted its pressure on the populations that possessed it and its antecedents" or "true language, via the emergence of syntax, was a catastrophic event, occurring within the first few generations of Homo sapiens" (Bickerton 1995), because "a relatively simple algorithmic basis of language" made a "language-ready brain" (Boeckx and Benitez-Burraco 2014) in which human-level language "can emerge virtually instantaneously."

We first dispose of the pure gradualist position.  Pinker and Bloom (Pinker and Bloom 1990) wrote that "Every detail of grammatical competence that we wish to ascribe to selection must have conferred a reproductive advantage on its speakers, and this advantage must be large enough to have become fixed in the ancestral population.  And there must be enough time and genomic space separating our species from nonlinguistic primate ancestors."  Perhaps they misspoke; it is extraordinarily well documented that a mutation may well become fixed in a population merely by not being selected against; far from being such a large advantage.  It also is well documented (as they mention) that a mutation without advantage may be yoked to another advantageous (but possibly unrelated) change (as in exaptations, or "spandrels").

In the present instance, brain size increase was far from gradual: the fossil record most typically contains either incredibly fast bursts of growth, or punctate jumps (both supporting and contrasting evidence has been forwarded; e.g., (Eldredge and Gould 1972; Gould 2002; Dawkins 1996).  The origin of these exceptional increases remains controversial, but the timeline itself is well documented.

Tattersall tells us that since "brain size increase was a property of the Homo clade in general," it "evidently had nothing to do with how modern Homo sapiens, specifically, did or does its cognitive business."  He arrives at this conclusion from a consideration of the existing archeological record: contrasts between presumed Homo sapiens and Neanderthal sites, on one hand; and between pre- and post-Middle Stone Age eras on the other, when evidence of complex cognition appears in the forms of long-distance transportation, hide preservation, pierced and ochre-stained shells, and related events.  It is notable that the estimates of these "earliest" complex behavioral phenomena have repeatedly been pushed back, with almost every new archaeological finding; one might be forgiven for somewhat discounting the current versions of those estimates and instead actually extrapolating from the timeline of further discoveries and their remarkable tendency to cause backward revisions of the estimates.  Some evidence suggests extremely small and localized early populations of Homo sapiens (or their similar antecedents: it is notable that no sequencing yet exists of, e.g., Cro-magnon and other Upper Paleolithic purported modern humans), perhaps entailing a combination of changes in individuals and in societies.

All relevant ancestors are gone, and we do not know whether they had language.  We specifically propose that brains with neocortex-to-hippocampus ratio up to values of 0.9 may be characterized as equivalent to single unrestricted-stack automata, generating context free grammars; when the cortex-to-hippocampus ratio exceeds the value of 0.9, the cortical encephalization of what had previously been a purely hippocampal mechanism, generates the first nested stacks.  If there had been brain sizes intermediate to those of apes, on one hand, and Homo sapiens on the other, they might, according to this reasoning, have been constructing nested grammars (see Table 1).

| mm$^3$ | telenc | hipp | neo | neo/hipp |
|---|---|---|---|---|
| **tenrec** | 1,407 | 146 | 271 | 1.9 |
| **aye-aye** | 30,196 | 1,776 | 22,127 | 12.5 |
| **monk saki** | 24,920 | 834 | 21,028 | 25.2 |
| **baboon** | 154,987 | 3,398 | 140,142 | 41.2 |
| **macaque** | 71,080 | 1,353 | 63,482 | 46.9 |
| **gorilla** | 369,878 | 4,781 | 341,444 | 71.4 |
| **chimp** | 313,493 | 3,779 | 291,592 | 77.2 |
| **erectus** | 780,000 | 9,116 | 729,300 | 80.0 |
| **heidelb.** | 948,000 | 10,237 | 890,172 | 87.0 |
| **human** | 1,063,399 | 10,287 | 1,006,525 | **97.8** |

Table 1. Sizes of hippocampus, neocortex, and overall telencephalon, in a range of primates. The growth of these regions is commensurate with allometric predictions (see text); the slope of neocortical growth is far higher than that of hippocampal growth, causing the ratio of neocortex to hippocampus to grow precipitously, reaching nearly an absolute ceiling at its maximum in humans. (Tentative projected hypothetical figures based on calvaria are inserted for *H.erectus* and *H.heidelbergensis*.) (Data from Stephan et al.)

The findings presented herein appear to provide a candidate explanatory account of allometric evolution of human all-and-only capacities, including natural language: the apparent abrupt (qualitative) changes to behavioral and cognitive capabilities arise as a particular brain circuitry achieves specific quantitative thresholds of stacks in single-stack pushdown automata, as a direct consequence of allometrically growing cortical-subcortical ratios, independent of any effects of natural selection or other extrinsic influences.

This may have happened extraordinarily suddenly, by virtue of the allometric, yet enormous, change to cortical-hippocampal ratios, with little or no need for adaptation, as with all allometric steps.

The category of human all-and-only abilities are fundamental to humanness; in a crucial way, they quantitatively define what we intrinsically are.

Human minds are, by that evidence, nested-stack automata, intrinsically computing indexed grammars. Our natural languages clearly reflect this; further quantitative analyses of other abilities may demonstrate that this is also true of any capacity that is in all, and only, humans.

*Acknowledgments:* This work was supported in part by grants from the Office of Naval Research (ONR) and the Defense Advanced Research Projects Agency (DARPA). Many thanks to Tony Rodriguez and Eli Bowen for highly useful discussions.